\pdfoutput=1
\PassOptionsToPackage{table}{xcolor}
\documentclass[journal,twoside,web]{ieeecolor}
\usepackage{generic}
\usepackage[T1]{fontenc}
\usepackage[utf8]{inputenc}
\usepackage{cite}
\usepackage{microtype}
\usepackage{amsmath,amssymb,amsfonts}
\usepackage{graphicx}
\usepackage{booktabs}
\usepackage{tabularx}
\usepackage{threeparttable}
\usepackage{array}
\usepackage{pifont}
\usepackage{siunitx}
\usepackage{textcomp}
\usepackage{url}
\usepackage[hidelinks]{hyperref}
\usepackage{orcidlink}
\usepackage[nameinlink,capitalise]{cleveref}

\newif\ifreviewmarks
\IfFileExists{highlight-local.tex}{\reviewmarkstrue}{\reviewmarksfalse}
\ifreviewmarks
  \usepackage{soulutf8}
  \sethlcolor{yellow}
  \DeclareRobustCommand{\reviewbox}[1]{{\setlength{\fboxsep}{0pt}\colorbox{yellow}{#1}}}
  \newcommand{\reviewmath}[1]{\reviewbox{$#1$}}
  \newcommand{\reviewindex}[1]{\reviewbox{$\scriptstyle #1$}}
\else
  \DeclareRobustCommand{\hl}[1]{#1}
  \DeclareRobustCommand{\reviewbox}[1]{\ifmmode\mbox{#1}\else#1\fi}
  \newcommand{\reviewmath}[1]{\ensuremath{#1}}
  \newcommand{\reviewindex}[1]{#1}
\fi

\newcommand{\cmark}{\textcolor{green!70}{\ding{51}}}
\newcommand{\xmark}{\textcolor{red!80}{\ding{55}}}
\newcommand{\eg}{e.g.}
\newcommand{\ie}{i.e.}
\newcommand{\etal}{et al.}

\newcommand{\Rp}{{R^{\mathrm{(p)}}}}
\newcommand{\Rv}{{R^{\mathrm{(v)}}}}
\providecommand{\refname}{References}
\DeclareMathOperator{\Log}{Log}
\makeatletter
\providecommand{\l@en}{\l@english}
\providecommand{\l@eng}{\l@english}
\def\thebibliography#1{%
    \section*{\refname}%
    \addcontentsline{toc}{section}{\refname}%
    \footnotesize
    \vskip 0.3\baselineskip plus 0.1\baselineskip minus 0.1\baselineskip%
    \list{\@biblabel{\@arabic\c@enumiv}}{%
    \settowidth\labelwidth{\@biblabel{#1}}%
    \leftmargin\labelwidth
    \advance\leftmargin\labelsep\relax
    \itemsep 0pt plus 0pt\relax
    \usecounter{enumiv}%
    \let\p@enumiv\@empty
    \renewcommand\theenumiv{\@arabic\c@enumiv}}%
    \let\@IEEElatexbibitem\bibitem%
    \def\bibitem{\@IEEEbibitemprefix\@IEEElatexbibitem}%
    \def\newblock{\hskip .11em plus .33em minus .07em}%
    \if@technote
        \sloppy\clubpenalty4000\widowpenalty4000\interlinepenalty100%
    \else
        \sloppy\clubpenalty4000\widowpenalty4000\interlinepenalty500%
    \fi
    \sfcode`\.=1000\relax}

\def\bstctlcite{\@ifnextchar[{\@bstctlcite}{\@bstctlcite[@auxout]}}
\def\@bstctlcite[#1]#2{\@bsphack
  \@for\@citeb:=#2\do{%
    \edef\@citeb{\expandafter\@firstofone\@citeb}%
    \if@filesw\immediate\write\csname #1\endcsname{\string\citation{\@citeb}}\fi}%
  \@esphack}
\makeatother
\def\BibTeX{{\rm B\kern-.05em{\sc i\kern-.025em b}\kern-.08em
    T\kern-.1667em\lower.7ex\hbox{E}\kern-.125emX}}

\begin{document}
\bstctlcite{IEEEexample:BSTcontrol}

\title{Echo-POSED: Geometric Self-Distillation for Echocardiography Guidance}

\author{Elias~Stenhede
,
Edvart~Grüner~Bjerke
,
Joanna~Sulkowska
,
Eivind~Bjørkan~Orstad
,
Ole~Jakob~Elle
,
Ulysse~Côté-Allard
,
and Arian~Ranjbar
\thanks{This work has been submitted to the IEEE for possible publication. Copyright may be transferred without notice, after which this version may no longer be accessible.}
\thanks{E. Stenhede and E. G. Bjerke contributed equally.}
\thanks{U. Côté-Allard and A. Ranjbar contributed equally as senior authors.}
\thanks{E. Stenhede, A. Ranjbar, and E. B. Orstad are with Akershus University Hospital, Norway (e-mail: elias.stenhede@ahus.no).}
\thanks{E. G. Bjerke, J. Sulkowska, O. J. Elle and U. Côté-Allard are with the University of Oslo, Norway.}
\thanks{O. J. Elle is also with Oslo University Hospital, Norway.}}

\maketitle

\begin{abstract}
    Echocardiographic probe guidance typically relies on expert \hl{annotations} or tracked probe trajectories, \hl{which limits the use of routinely acquired data for training}. We introduce Echo-POSED \hl{(Echocardiography POse learning via SElf-Distillation)}, a self-supervised framework for real-time \hl{echocardiography guidance toward apical views} that recommends probe adjustments directly from two-dimensional ultrasound images. During training, two-dimensional views are resliced from three-dimensional videos under known probe perturbations, and geometric self-distillation enforces equivariance to probe motion while maintaining invariance to cardiac phase. Trained on the EchoXFlow dataset, Echo-POSED learns a stable anatomical representation across patients and maintains geometric consistency under virtual probe perturbations on held-out data and four external public datasets spanning multiple vendors. \hl{On the held-out data}, the mean absolute \hl{error during a rotation-only sweep about} the beam axis was \reviewbox{\ang{8.6}}. \hl{In simulated guidance against a method trained on expert annotations, Echo-POSED was better on two datasets, worse on one, and not significantly different on two. Across all 3,015 attempts on the external datasets, two view classifiers recognised the requested view in 72--77\%.} We publicly release our model code, model weights, and the manual annotations we created for evaluation. \url{https://github.com/Ahus-AIM/Echo-POSED}

\end{abstract}


\section{Introduction} \label{sec:intro}
Transthoracic echocardiography (TTE) is the primary imaging modality in cardiology, providing real-time visualisation of cardiac structure and function~\cite{langRecommendationsCardiacChamber2015}. Its clinical utility, however, hinges on acquiring reproducible, standardised 2D views that enable clinically actionable measurements~\cite{hagendorffValidReproducibleQuantitative2023}. Obtaining such views requires precise probe positioning and angulation, a process that is highly operator-dependent and demands substantial training and experience~\cite{mitchellGuidelinesPerformingComprehensive2019, kirkpatrickRecommendationsEchocardiographyLaboratories2020}. When such expertise is lacking, suboptimal views might lead to incomplete assessment or misdiagnosis. Since the demand for echocardiographic examinations far exceeds the availability of trained specialists, there is a growing effort to automate acquisition guidance, both to support novice operators and to enable examinations when specialist expertise is unavailable. Another potential application of such systems is fully autonomous robotic ultrasound systems~\cite{biMachineLearningRobotic2024}. Yet most existing guidance pipelines still depend on explicit supervision, such as tracked probe motion, or manually defined targets such as labelled planes~\cite{yue_echoworld_2025, pasdeloupRealTimeEchocardiographyGuidance2023, ferrazAssistedProbeGuidance2023}. Using volumetric data to train a self-supervised pose regressor has been explored for intraoperative X-ray registration~\cite{gopalakrishnanIntraoperative2D3D2024}; however, the model is trained separately for each patient, and generalisation across patients is not addressed. A challenge is to reduce reliance on specialised supervision so that guidance can be learned from the much larger pool of routinely acquired echocardiographic data while generalising across patients.


\subsection{Problem formulation}  \label{sec:problem-formulation}
\begin{figure}[ht]
    \centering
    \includegraphics[width=\linewidth]{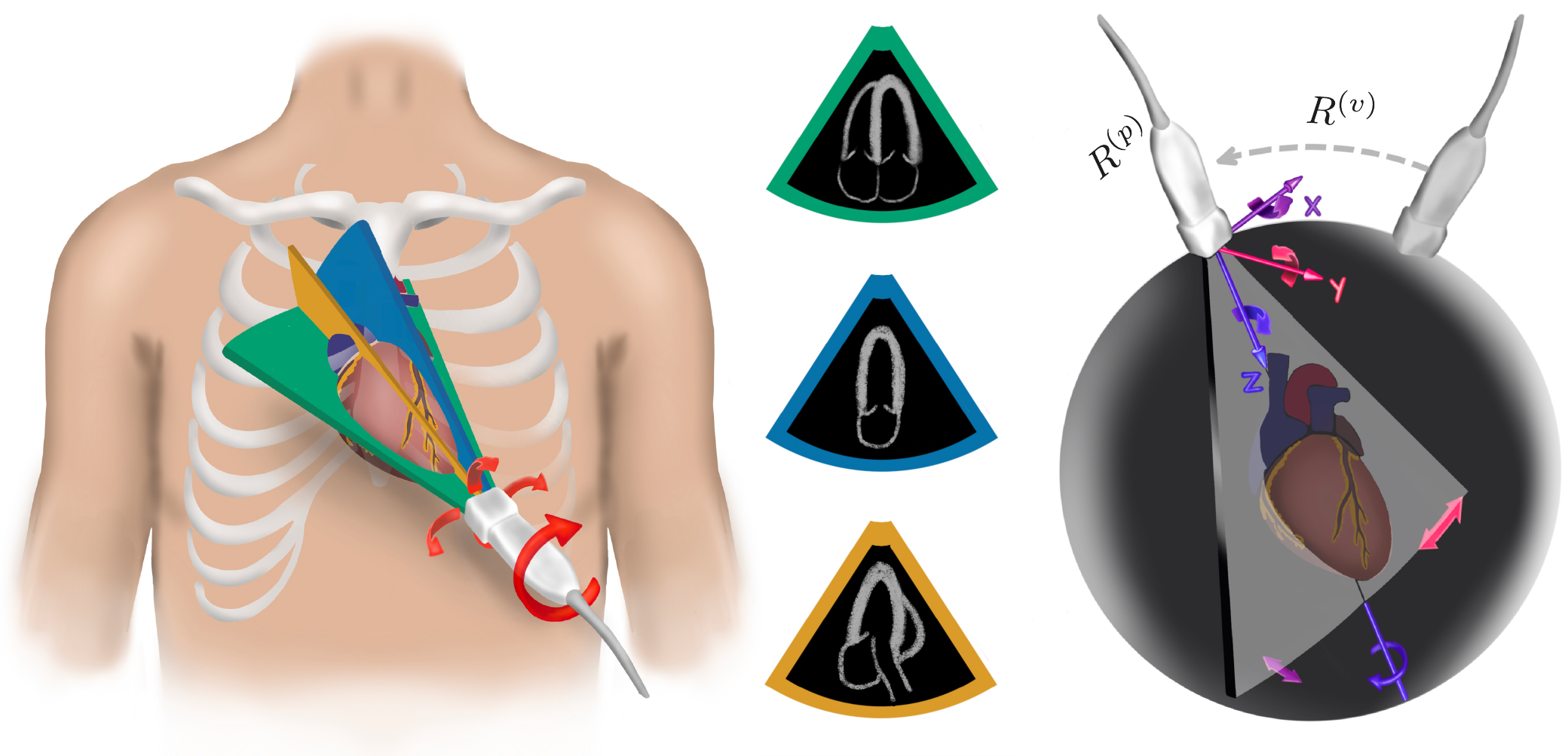}
    \caption{Echocardiographic acquisition setup and the effect of probe motion on the observed 2D image, with examples of three standard views. We parametrised the probe pose by two rotations: $\Rv$ for the probe motion on a sphere, and $\Rp$ for its orientation.}
    \label{fig:prob-form}
\end{figure}

We consider the problem of guiding an ultrasound probe toward \hl{a target apical view} in real time.
Ultrasound probe motion can be described by a rigid transformation, \ie, a 3D rotation and a 3D translation. Because the probe must remain in contact with the patient’s chest, this reduces to a problem with five degrees of freedom~\cite{pasdeloupRealTimeEchocardiographyGuidance2023}. We restrict translation to a sphere $\mathrm{S}^2$, \hl{centred on the central beam at half the imaged depth}; see \Cref{fig:prob-form}. This space is minimally represented by $\mathrm{SO}(3) \times \mathrm{S}^2$. To maintain a consistent Lie group structure and allow symmetric representations, we lift this space to
\begin{equation}\label{eq:group}
    G = \mathrm{SO}(3) \times \mathrm{SO}(3).
\end{equation}
This formulation stands in contrast to previous work that models probe pose using rigid motion~\cite{yue_echoworld_2025, huhAIDrivenViewGuidance2025} or assumes correct probe placement and considers rotation in isolation~\cite{pasdeloupRealTimeEchocardiographyGuidance2023}. Our modelling choice is further explained in \Cref{sec:representation-of-rotations}.

\subsection{Contributions}
\begin{enumerate}
    \item We introduce a geometric self-distillation framework for echocardiographic view guidance that requires neither manual annotation nor robotic trajectory data.
    \item We demonstrate Echo-POSED's ability to perform guidance toward apical views across public 3D echocardiography datasets.
    \item We present a benchmark on open data for evaluating echocardiography guidance and pose estimation.
\end{enumerate}
\section{Related Work}
Earlier deep learning work has established performance in multi-class echocardiographic view recognition~\cite{madaniFastAccurateView2018, ostvikRealTimeStandardView2019} and real-time quality assessment. Quality-focused methods have also been used to estimate acquisition errors, such as apical foreshortening~\cite{kimAutomatedDetectionApical2023, smistadRealTimeAutomaticEjection2020}. However, these produce either a discrete class label or a quality score, neither of which encodes the relationship between the current and target probe pose needed to guide the operator toward the desired view.

Probe guidance has been modelled as a sequential decision problem learned from expert demonstrations~\cite{narangUtilityDeepLearningAlgorithm2021, baoRealWorldVisualNavigation2024, jiangStructureawareWorldModel2025}. These methods typically require specialised acquisition setups that track probe motion, yielding temporally matched 2D frames and probe trajectories. Recent work has extended this to motion-aware world models~\cite{yue_echoworld_2025}. Such pipelines depend on expensive, specialised data collection and explicit identification of all standard views and their timestamps within each examination. 

Most closely related to our setting are methods that use 3D echocardiography to extract 2D slices with known differences in virtual probe pose. Pasdeloup \etal~\cite{pasdeloupRealTimeEchocardiographyGuidance2023} demonstrate real-time standard-view extraction of long-axis images using this method. While effective in real-world settings~\cite{saboRealtimeGuidingDeep2023, saboRealtimeGuidanceDeep2023}, such approaches depend on dense expert annotation, and therefore do not scale to large unlabelled clinical archives.

A related guidance objective has been pursued in fetal ultrasound, where several works learn pose regression from slices extracted from 3D volumes~\cite{liStandardPlaneLocalisation2018, liStandardPlaneDetection2018, menScanAheadSimplifyingStandard2025, linguraru_pose-guidenet_2024, droste_automatic_2020}. Among these, Yeung~\etal~\cite{yeungLearningMap2D2021} take a step toward self-supervision by generating training pairs through reslicing, but manual alignment of each training volume is still required.

In contrast to imitation methods and annotation-heavy slicing pipelines, \hl{Echo-POSED learns} pose representations directly from routinely acquired 3D videos, requiring neither manual annotation nor paired probe trajectories. 

\section{Methods}
This work employs geometric self-distillation~\cite{midtvedtSingleshotSelfsupervisedObject2022} to learn a guidance system that is equivariant to the group $G$ defined in \Cref{eq:group}. Intuitively, equivariance provides the structure needed for probe guidance: even if the \emph{optimal angle} required to obtain a specific echocardiographic view is unknown, changes of the orientation of the patient's heart induce corresponding changes in the optimal probe orientation.

We let $T_g$ denote the action of $g \in G$ on the input image space, corresponding to physical rotation and translation of the ultrasound probe, which induces changes to the observed echo image. Equivariance of $f_\theta:\mathbb{R}^{\text{H}\times\text{W}}\to G$ at some input $x\in \mathbb{R}^{\text{H}\times\text{W}}$ is expressed as
\begin{equation}\label{eq:equivariance}
    f_\theta\!\left(T_g x\right)
    =
    f_\theta(x)\, \reviewmath{g},
    \quad \forall g \in G .
\end{equation}
Importantly, we do not assume the existence of a globally consistent coordinate system across training volumes. We therefore seek to learn pose representations up to a global gauge transformation, relying on relative consistency \hl{during training, without requiring a predefined global reference frame}.

\subsection{Training objective}
\label{sec:training-objective}
As $G$ is a group, \Cref{eq:equivariance} can be rewritten as 
\begin{equation}\label{eq:equivariance-rewrite}
    f_\theta\!\left(T_g x\right) \reviewmath{g^{-1}}
    =
    f_\theta(x),
    \quad \forall g \in G,
\end{equation}
and in particular, for any $g_1, g_2 \in G$,
\begin{equation}\label{eq:equivariance-rewrite-ex}
    f_\theta\!\left(T_{g_1} x\right) \reviewmath{g_1^{-1}}
    =
    f_\theta\!\left(T_{g_2} x\right) \reviewmath{g_2^{-1}}.
\end{equation}
This expresses a consistency constraint: the model $f_\theta$ must preserve the difference between views. More formally, the training objective enforces consistency of the network output under transformations from the group $G$, as implied by equivariance. In practice, this is achieved by comparing the predictions obtained from multiple transformed observations of the same underlying anatomy. 

\subsubsection{Spatial equivariance and temporal invariance}\label{eq:equivariance-loss}
The left ventricle undergoes a twisting motion during the cardiac cycle~\cite{nakataniLeftVentricularRotation2011}, but the ultrasound probe should remain fixed when a correct view is obtained. Consequently, a guidance system should produce recommendations that are invariant to the cardiac phase. 

\begin{figure*}[!t]
  \centering
  \makebox[\linewidth][c]{\includegraphics[width=\linewidth]{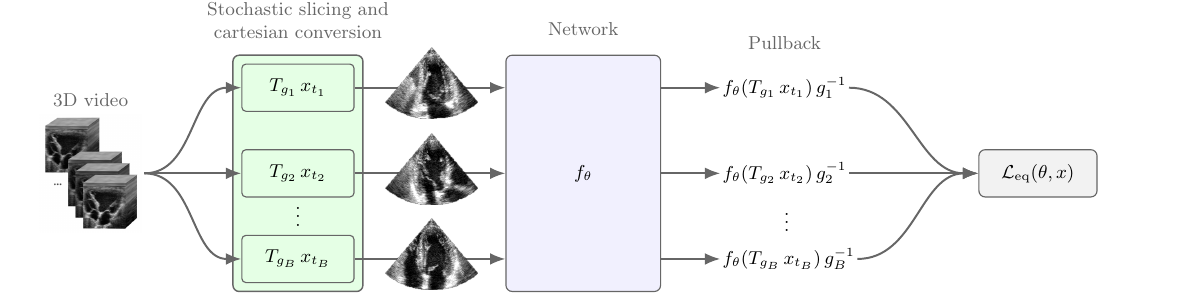}}
  \caption{Three-dimensional echocardiography videos are used to generate a batch of two-dimensional frames with known spatiotemporal differences. The network is trained to ignore temporal differences but be equivariant to spatial transformations.}
  \label{fig:ensemble}
\end{figure*}
In each training iteration, we sample a single echocardiographic 3D video $X=(X_1,\dots,X_T)$ and draw $B$ independent transformation–time pairs $\{(g_i,t_i)\}_{i=1}^B$, where $g_i \sim \mathcal{P}_G$ and $t_i \sim \mathrm{Unif}(\{1,\dots,T\})$. A suitable choice for the distribution $\mathcal{P}_G$ is supported on rotations and translations in the vicinity of the left ventricular axis, since equivariance requires the presence of anatomical landmarks (see \Cref{sec:setup} for our choice of $\mathcal{P}_G$). We let $T_{g} x_{t}$ denote a 2D slice generated from the 3D volume $X_t$ at a probe pose defined by $g$. We enforce spatial equivariance and temporal invariance using the following pairwise loss
\begin{equation}\label{eq:emp-risk-time}
    \mathcal{L}_\text{eq}(\theta,x)
    =
    \operatorname*{mean}_{i\neq j}
    d\!\left(
    f_\theta(T_{g_i} x_{t_i})\, \reviewmath{g_i^{-1}},
    f_\theta(T_{g_j} x_{t_j})\, \reviewmath{g_j^{-1}}
    \right),
\end{equation}
where $d$ is the geodesic distance and the mean is over ordered pairs. The batch generation process is illustrated in \Cref{fig:ensemble}. Note that the loss only concerns differences between views; \emph{the choice of reference frame is still arbitrary} at this point. The following sections address how to establish a reference frame, and how to represent rotations. 

\subsubsection{Aligning the ensemble}
To improve prediction accuracy and estimate uncertainty, we independently train an ensemble of $K$ models. Because the objective in \Cref{eq:emp-risk-time} determines the learned representation only up to a global gauge transformation, different ensemble members may adopt different coordinate frames. Their predictions therefore cannot be directly combined.

\begin{figure*}[t]
  \centering
  \makebox[\linewidth][c]{\includegraphics[width=\linewidth]{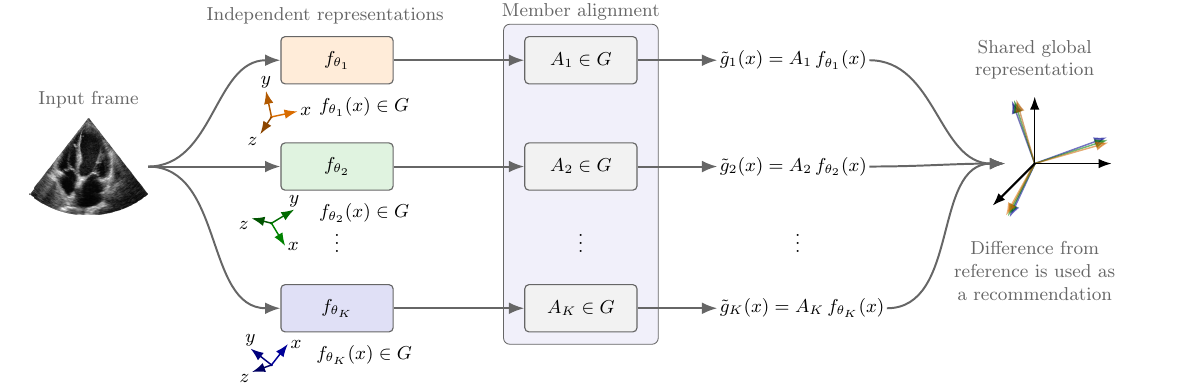}}
  \caption{Member alignment ensures a shared global representation by fitting an adapter for each ensemble member. The difference between the mean prediction in the shared global representation and a known reference is used as a guidance signal for the user. The ensemble disagreement is used as an uncertainty score.}
  \label{fig:gauge}
\end{figure*}

We therefore equip the ensemble with \textit{member alignment}: \hl{for} \reviewmath{k=1,\dots,K}, model $k$ predicts $f_{\theta_k}(x)\in G$ and carries a left-multiplicative adapter $A_k\in G$, so that its aligned prediction is
\begin{equation}\label{eq:member-alignment}
    \tilde g_k(x)=A_kf_{\theta_k}(x).
\end{equation}
The adapters are the coordinate transformations between the output spaces the members learned independently. \hl{We fix every member's gauge against a common external reference}: the rotations applied when slicing a calibration set $X_{\mathrm{cal}}$. \hl{The choice of calibration set only determines this common frame.} Writing $g_i\in G$ for the rotation applied to produce $x_i$, each adapter is the orthogonal-Procrustes solution
\begin{equation}\label{eq:gauge-fit}
    A_k^\ast
    =
    \underset{A \in G}{\operatorname{argmin}}
    \sum_{x_i \in X_{\mathrm{cal}}}
    \left\lVert
        A f_{\theta_k}(x_i) - g_i
    \right\rVert_{\mathrm{F}}^{2},
\end{equation}
which is obtained in closed form from the singular value decomposition of $\sum_i g_i f_{\theta_k}(x_i)^{\top}$. An adapter is fitted this way for each factor of every member on the development set, and every experiment uses the aligned predictions. We denote the projected mean of the aligned ensemble predictions as $F_\Theta(x)$; details on how the mean value is calculated are given in \Cref{sec:ensemble-predictions-and-uncertainty} and illustrated in \Cref{fig:gauge}.
\subsection{Representation of rotations and translations}\label{sec:representation-of-rotations}
We model the probe using two rotations rather than an explicit rotation-translation pair. We adopt this design for three reasons.
\begin{enumerate}
  \item \textbf{Slice generation.} During training, 2D views are extracted from 3D volumes. Constraining motion to spheres around \hl{a pivot inside the volume} prevents sampling planes from drifting outside the anatomy.
  \item \textbf{Unit-free parametrisation.} Using two rotations avoids choosing a physical translation scale (\eg, centimetres) and instead represents \hl{probe translation} in the same Lie-group framework as probe rotation.
  \item \textbf{Scale-invariance.} Left ventricular foreshortening and related appearance changes are captured as angular effects, reducing dependence on absolute heart size.
\end{enumerate}
Rotations in 3D are elements of $\mathrm{SO}(3)$, typically represented as $3\times3$ orthogonal matrices with positive determinant. For gradient-based learning, the choice of parametrisation is critical because it affects the stability of optimisation~\cite{zhou_continuity_2019}. Common $\mathrm{SO}(3)$ parametrisations are listed in \Cref{tab:so3_representations}.
Previous work in echocardiographic guidance has mainly used Euler angles~\cite{yue_echoworld_2025} or discrete orientation classes~\cite{pasdeloupRealTimeEchocardiographyGuidance2023}. In contrast, we predict the probe rotation and translation ($\Rp$ and $\Rv$) as unconstrained $3\times3$ matrices projected to $\mathrm{SO}(3)$ via SVD. The choice is motivated empirically in our setting, and avoids the discontinuities associated with representations embedded in fewer than five dimensions~\cite{levinson_analysis_2020, zhou_continuity_2019}.
\begin{table}[t]
    \centering
    \begin{threeparttable}
    \footnotesize
    \setlength{\tabcolsep}{2pt}
    \caption{Common representations of $\mathrm{SO}(3)$ rotations~\cite{geist_learning_2024}.}
    \label{tab:so3_representations}
    \begin{tabular*}{\linewidth}{@{\extracolsep{\fill}}lrcc@{}}
        \toprule
        Representation &
        Dim. &
        Domain &
        \hl{Continuous} \\
        \midrule
        Euler & 3 & $\mathbb{R}^3$ & \xmark \\
        Exponential coordinates & 3 & $\mathbb{R}^3$ & \xmark \\
        Unit quaternions & 4 & $S^3$ & \xmark \\
        Axis--angle & 4 & \reviewmath{S^2\times[0,\pi]} & \xmark \\
        6D + \hl{Gram--Schmidt} & 6 & $\mathbb{R}^{3\times 2}$ & \cmark \\
        9D + \hl{singular value decomposition} & 9 & $\mathbb{R}^{3\times 3}$ & \cmark \\
        \bottomrule
    \end{tabular*}
    \end{threeparttable}
\end{table}

\subsubsection{Unobservable rotation} We model the system using $\mathrm{SO}(3)\times\mathrm{SO}(3)$, with one rotation for \hl{probe translation and one for probe rotation}. This introduces a gauge ambiguity: applying the same rotation about the $z$-axis (see \Cref{fig:prob-form} for coordinate conventions) to both factors leaves the extracted 2D slice unchanged. Consequently, different elements of $G$ can correspond to the same observation. The distance $d$ in \Cref{eq:emp-risk-time} should thus not penalise differences that are not observable in the 2D image domain. We resolve the ambiguity by having the network apply any rotation about the $z$-axis to the probe only, reflecting that rotating a probe is easier than rotating a patient. In practice, this is implemented via the dataloader. \hl{Equivariance is therefore enforced only for these representatives.}

\hl{Each slice is generated by orienting the probe with} \reviewmath{R^{(\mathrm{p})}}\hl{ and then applying the movement} \reviewmath{R^{(\mathrm{v})}}\hl{, defined in the oriented probe's frame, which rotates the sampling plane and its apex about a pivot on the central beam at half the imaged depth. The loss and uncertainty are computed directly on these two rotations.}

\subsection{Ensemble prediction and uncertainty}\label{sec:ensemble-predictions-and-uncertainty}
At inference time, the guidance signal is obtained by aggregating the aligned predictions of the \(K\) ensemble members, as illustrated in \Cref{fig:gauge}. We denote the decomposition of group elements by rotation around the probe origin $R^{\mathrm{(p)}}$ and \hl{the volume pivot} $R^{\mathrm{(v)}}$ as $g=(R^{\mathrm{(v)}}, R^{\mathrm{(p)}})\in G$. For an input frame \(x\), member \(k\) produces an aligned prediction $\tilde g_k(x)=A_k f_{\theta_k}(x) = \big(\tilde R_k^{\mathrm{(v)}}(x),\tilde R_k^{\mathrm{(p)}}(x)\big)\in G$. We compute the projected (extrinsic) mean on each factor \reviewmath{c\in\{\mathrm{v},\mathrm{p}\}} as
\begin{equation}
\bar R^{(\reviewindex{c})}(x)
=
\operatorname{Proj}_{\mathrm{SO}(3)}
\left(
\frac{1}{K}\sum_{k=1}^K \tilde R_k^{(\reviewindex{c})}(x)
\right),
\label{eq:ensemble_mean_product}
\end{equation}
where $\operatorname{Proj}_{\mathrm{SO}(3)}$ denotes orthogonalisation (SVD projection) to the closest rotation matrix in Frobenius norm. The ensemble mean is
\begin{equation}
\label{eq:ens-mean}
F_\Theta(x)=\big(\bar R^{\mathrm{(v)}}(x),\bar R^{\mathrm{(p)}}(x)\big).
\end{equation}
In addition to a point estimate, we report an uncertainty score based on ensemble disagreement. We measure mean squared disagreement in the tangent space at $F_\Theta(x)$ using the component-wise logarithm, \hl{with the frame argument implicit in the rotations},
\begin{equation}
u_\text{ens}(x)
=
\sqrt{\frac{1}{2K}\sum_{k=1}^K
\sum_{\reviewindex{c}}
\left\|
\Log\!\left((\bar R^{(\reviewindex{c})})^\top \tilde R_k^{(\reviewindex{c})}\right)
\right\|_2^2}.
\label{eq:ensemble_var_product}
\end{equation}
This is the \hl{square root of the} sample mean of squared geodesic distances under the product metric between individual members and the ensemble mean. 
\section{Data sources}
We utilise one dataset for training and internal validation, and test our method on, to our knowledge, all publicly available 3D TTE datasets. For training, we utilise EchoXFlow, a public clinical beamspace echocardiography dataset that preserves native acquisition geometry~\cite{stenhedeEchoXFlowBeamspaceEchocardiography2026}. \hl{EchoXFlow includes 716 3D+t B-mode recordings from 487 examinations.} Key characteristics of EchoXFlow are summarised in \Cref{tab:echo_datasets}, together with the other 3D TTE datasets that are published at the time of writing~\cite{bernard_standardized_2016, tobon-gomezBenchmarkingFrameworkMyocardial2013, zhaoMITEADatasetMachine2022, vukadinovicAutomatedInterpretable2D2025}.
\begin{table}[t]
    \centering
    \begin{threeparttable}
    \footnotesize
    \setlength{\tabcolsep}{1pt}
    \caption{Publicly available 3D echocardiography datasets.}
    \label{tab:echo_datasets}
    \begin{tabular*}{\linewidth}{@{\extracolsep{\fill}}lrll@{}}
        \toprule
        Dataset &
        \hl{Dataset size} &
        Geometry &
        Vendor \\
        \midrule
        EchoXFlow~\cite{stenhedeEchoXFlowBeamspaceEchocardiography2026} & \hl{487 examinations} & Beamspace, time-resolved & GE \\
        CETUS~\cite{bernard_standardized_2016} & 45 \hl{individuals} & Cartesian & GE/PHI/SIE  \\
        EchoNet3D~\cite{vukadinovicAutomatedInterpretable2D2025} & 4 \hl{individuals} & Beamspace, time-resolved & PHI \\
        MITEA~\cite{zhaoMITEADatasetMachine2022} & 134 \hl{individuals} & Cartesian & SIE \\
        STACOM~\cite{tobon-gomezBenchmarkingFrameworkMyocardial2013} & 15 \hl{individuals} & Cartesian & PHI \\
        \bottomrule
    \end{tabular*}
    \begin{tablenotes}
        \scriptsize
        \item[] GE = General Electric; PHI = Philips; SIE = Siemens.
    \end{tablenotes}
    \end{threeparttable}
\end{table}

\subsection{Co-registered 2D--3D view pairs}\label{sec:coregistered-pairs}
For evaluation, we use a set of co-registered 2D--3D pairs derived from a held-out portion of EchoXFlow. Each pair couples a \hl{native} 2D apical recording with a pose in a 3D volume from the same examination, such that reslicing the volume at that pose reproduces the imaging plane of the 2D recording. Every pair carries an expert-confirmed view label, one of A4C, A2C, or A3C. \hl{After ECG-based temporal matching, planes with the highest normalised cross-correlation, computed with fast Fourier transforms in beamspace, were proposed and then approved manually by two of the authors. Every retained pair also passed a classifier screen in which EchoPrime}~\reviewbox{\cite{vukadinovicComprehensiveEchocardiogramEvaluation2026}}\hl{ and EchoFine}~\reviewbox{\cite{naidooRobustFinegrainedEchocardiographic2026}}\hl{ both assigned the stated view to the native 2D frame.} Because 3D acquisition often uses a shallower sector than 2D, we retain only pairs in which both the 2D and the 3D sector extend beyond 12\,cm, \hl{to reduce the risk of truncating the left ventricle}. This yields 171 pairs from \hl{61 examinations in 58 individuals}, at most one per examination and view: 56 A2C, 57 A3C, and 58 A4C. \hl{We use this depth-filtered set for the paired-cine analysis, the guidance simulations on EchoXFlow, and the external consensus targets.}

\section{Model implementation}\label{sec:setup}
The model architecture was designed for real-time CPU inference and reflects many clinical settings. Each ensemble member is an ImageNet-initialised EfficientNet-B0~\cite{tan_efficientnet_2019} operating on single-channel $224\times224$ scan-converted slices, with the classification head replaced by a linear head emitting 18 values: nine unconstrained values per factor, each projected to $\mathrm{SO}(3)$ by SVD; this gives 4.0\,M parameters per member. The SVD projection to $\mathrm{SO}(3)$ acts on $3\times3$ matrices and is therefore negligible in cost.

\hl{The four-member ensemble requires 3.1 GFLOPs per frame. Median latency from a native beamspace frame to the correction is 44 ms on a Threadripper PRO 9965WX CPU with eight threads and 9 ms on an RTX 5090 GPU.}

\subsection{Optimisation and cross-validation}
An ensemble comprising 4 models was trained on a single RTX 5090 for 20 epochs with the Muon optimiser for weights with dim $\geq 2$ and AdamW for biases and normalisation layers~\cite{loshchilovDecoupledWeightDecay2019, jordanMuonOptimizerHidden2024}. The learning rate was set to $0.0037$ with a cosine schedule down to $0.000037$ for both optimisers, and no weight decay was used. For each batch, translations and rotations were sampled to emulate realistic probe perturbations around the left ventricular axis. Probe rotations were drawn uniformly about the $z$-axis (\ang{0} to \ang{360}) and from \ang{-10} to \ang{10} around the $x$- and $y$-axes (see \Cref{fig:prob-form} for axis conventions used). Probe translations were realised as tilts about \hl{the pivot at half the imaged depth}: each tilt was a rotation about a random axis in the probe-local $x$--$y$ plane, with magnitude drawn uniformly up to \ang{10}. Slices $T_{g_i}x_{t_i}$ were generated on-the-fly from a 3D video $X$. Half of the batch was sampled independently, after which horizontal mirroring was applied with the corresponding transformation update, yielding a $\times 2$ effective batch expansion at negligible cost. \hl{A mirrored slice equals the slice obtained after rotating the probe by 180° about the beam axis. The probe target is updated by this rotation and the translation tilts change sign.} Because training was data-loading-bound, ensemble training added minimal overhead. \hl{Each member was assigned one of the four development folds, which it used for validation while updating its weights on the other three.}

\subsection{Data cleaning, preprocessing, and augmentation}
Because 3D echo acquires volumetric data, maintaining adequate frame rate may require a narrower acquisition sector or lower spatial resolution than in 2D imaging~\cite{langEAEASERecommendations2012, mitchellGuidelinesPerformingComprehensive2019}. To ensure that our training data resembled 2D, which is the intended application, we included 3D volumes whose azimuth and elevation sector widths both reached at least 1 radian, \ang{57}, whose sector started within 1\,cm of the probe and extended to at least 10\,cm, and which contained at least 20 frames, yielding 433 3D videos. These were split, at the patient level, into development and test sets, following the five-fold split distributed with EchoXFlow. In total, 357 3D videos were used for model development, and 76 3D videos were held out for testing. Augmentation was applied partly in beamspace, before scan conversion: random piecewise-linear greyscale remapping, random reduction of the sector width with a soft edge, and random anisotropic smoothing to prevent the model from learning 3D-echocardiography-specific voxel resolution patterns. After scan conversion, we applied random gamma and blur perturbations and a random corner-sector mask to avoid the model learning to attend to the sector shape.

\section{Experiments}
To enable reproducible comparison in echocardiography guidance, we establish a public benchmark by evaluating Echo-POSED on publicly available datasets. We evaluate Echo-POSED on a patient-level held-out portion of EchoXFlow and on the external datasets listed in \Cref{tab:echo_datasets},  in three main ways:
\begin{enumerate}
    \item Raw angular accuracy and uncertainty signal via controlled \hl{probe-rotation and probe-translation} sweeps, reporting mean absolute angular error per sweep dimension, and conditioned on ensemble uncertainty $u_\text{ens}$.
    \item Anatomical correspondence, \hl{by comparing pose predictions for native 2D cines and their 3D reslices with each other and with the annotated planes}. 
    \item Closed-loop guidance, using the same view classifiers to score whether guidance reaches the intended standard view.
\end{enumerate}
\hl{Neither view classifier was used to train Echo-POSED. EchoPrime}~\reviewbox{\cite{vukadinovicComprehensiveEchocardiogramEvaluation2026}}\hl{ (11 views) and EchoFine}~\reviewbox{\cite{naidooRobustFinegrainedEchocardiographic2026}}\hl{ (47 classes) each score a single} \reviewmath{224\times224}\hl{ frame. We sum their probabilities into A2C, A3C, A4C, and Other and take the largest.}
\subsection{Angular regression within volumes}\label{sec:ang-regression}
To assess \hl{the angular response to rotations about the probe axes}, we extract 2D slices with known probe perturbations from 3D volumes. For each volume, we generate rotated slices about the probe axes without translation; axis conventions are defined in \Cref{fig:prob-form}. We apply a full \ang{360} sweep about the probe $z$-axis, together with independent \ang{-10} to \ang{10} sweeps about the probe $x$- and $y$-axes. The two-degree-of-freedom \hl{probe translation} $\Rv$ (\Cref{sec:representation-of-rotations}) is swept separately\hl{, so that errors can be attributed to each rotation,} holding $\Rp$ fixed and varying \hl{its} tilt about $x$ and $y$; both sweeps are summarised together in \Cref{tab:scatter-summarised}.
The model operates on single-frame inputs with no test-time augmentation. Predictions are obtained from the ensemble mean and uncertainty computed according to \Cref{eq:ens-mean,eq:ensemble_var_product}. Mean Absolute Angular Error (MAAE) is presented, and a normalised variant using permutation testing so that random performance equals 1.0. \hl{Dataset-specific preprocessing is available in the released evaluation code.}

\begin{figure*}[t]
    \centering
    \includegraphics[width=\linewidth]{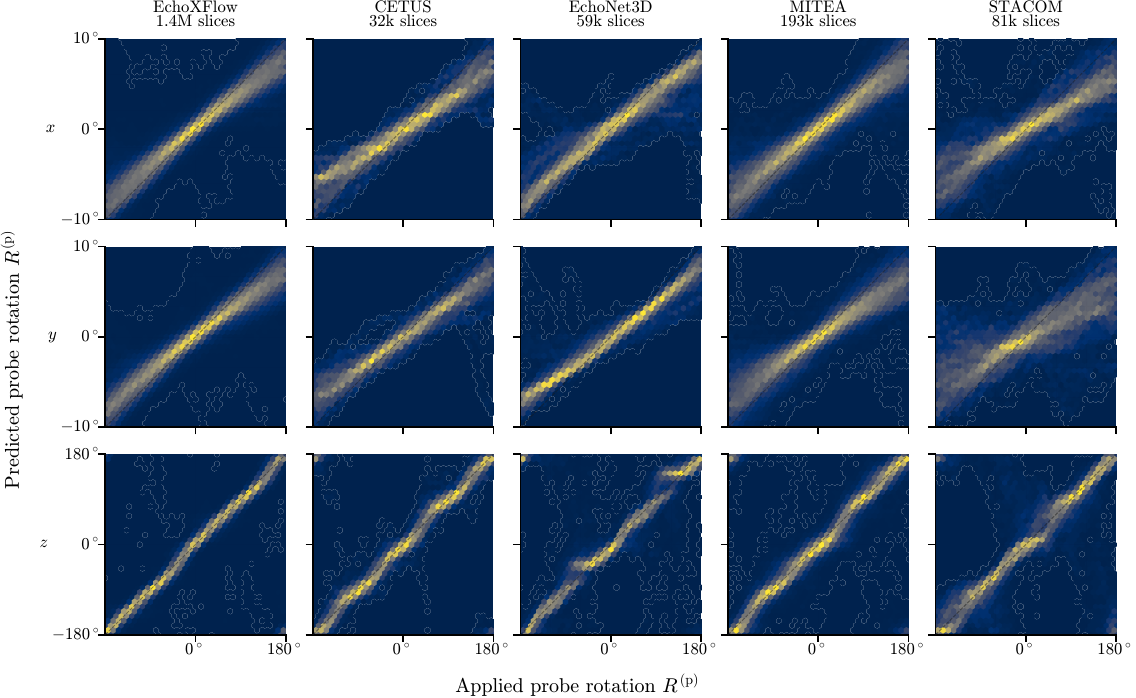}
    \caption{Model-predicted angle by applied angle across the held-out EchoXFlow fold and external test datasets. Rows correspond to rotations applied around the probe $x$-, $y$-, and $z$-axes without translation; each sweep comprises 360 applied angles per frame, and each frame's predictions are centred by removing their mean residual. Colour denotes bin count.}
    \label{fig:scatter-combined}
\end{figure*}
\Cref{fig:scatter-combined} presents calibration behaviour for rotations around the probe $x$-, $y$-, and $z$-axes across the independent echocardiography datasets. Rotations are applied to the virtual probe without translation and compared with the predicted rotations; points near the diagonal indicate geometric consistency. \hl{The EchoNet3D sweeps were generated from four individuals, so the many rendered slices represent repeated observations within a small cohort.}

Auxiliary angular-error summaries are provided in \Cref{tab:scatter-summarised}, where the datasets are sorted by the ensemble uncertainty $u_\text{ens}$ (\Cref{eq:ensemble_var_product}), computed per frame over both factors and averaged over all controlled sweeps. The best performance is observed on the held-out part of EchoXFlow, and the model further generalises to the other datasets, which were recorded using hardware from different vendors.

\begin{figure}[!tb]
    \centering
    \includegraphics[width=\linewidth]{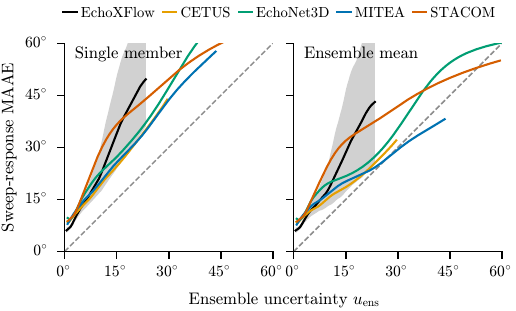}
    \caption{Realised $z$ sweep-response MAAE against the ensemble disagreement $u_\mathrm{ens}$, for a single ensemble member (left) and the ensemble mean (right). Curves are kernel-regression conditional means (Nadaraya--Watson using Silverman's rule of thumb in $\log u_\mathrm{ens}$), with EchoXFlow 95\% CI estimated by a 300-replicate cluster bootstrap band over recordings. Realised error \hl{rises} with disagreement, and the ensemble mean improves on the single member, especially at high uncertainty.}
    \label{fig:experiment1-uncertainty-calibration}
\end{figure}

\begin{table*}[t]
    \caption{Angular errors for both prediction heads. The probe-rotation panel applies a full rotation around the $z$-axis and rotations from $-10^\circ$ to $10^\circ$ around $x$ and $y$; the probe-translation panel fixes the probe rotation and sweeps \hl{its} two degrees of freedom $\Rv_x$ and $\Rv_y$. A normalised MAAE of $1$ represents random performance. The ensemble uncertainty $u_\mathrm{ens}$ (\Cref{eq:ensemble_var_product}) is a single per-frame quantity over both factors, averaged over all sweep renderings from both panels.}
    \label{tab:scatter-summarised}
    \centering
    \footnotesize
    \begin{tabular*}{\linewidth}{@{\extracolsep{\fill}}lrrrrrrrrrrr@{}}
    \toprule
     & \multicolumn{6}{c}{Probe rotation $\Rp$} & \multicolumn{4}{c}{Probe translation $\Rv$} & \\
    \cmidrule(lr){2-7}\cmidrule(lr){8-11}
     & \multicolumn{3}{c}{MAAE} & \multicolumn{3}{c}{nMAAE} & \multicolumn{2}{c}{MAAE} & \multicolumn{2}{c}{nMAAE} & \\
    \cmidrule(lr){2-4}\cmidrule(lr){5-7}\cmidrule(lr){8-9}\cmidrule(lr){10-11}
    Dataset & $x$ & $y$ & $z$ & $x$ & $y$ & $z$ & $x$ & $y$ & $x$ & $y$ & $u_\mathrm{ens}$ \\
    \midrule
    EchoXFlow & \reviewbox{\ang{1.3}} & \ang{1.4} & \reviewbox{\ang{8.6}} & \reviewbox{0.21} & 0.23 & 0.10 & \reviewbox{\ang{1.7}} & \ang{0.3} & \reviewbox{0.27} & \reviewbox{0.04} & \reviewbox{\ang{2.7}} \\
    CETUS & \reviewbox{\ang{1.7}} & \reviewbox{\ang{1.7}} & \reviewbox{\ang{11.3}} & \reviewbox{0.30} & \reviewbox{0.29} & \reviewbox{0.13} & \reviewbox{\ang{4.1}} & \reviewbox{\ang{0.7}} & \reviewbox{0.77} & \reviewbox{0.10} & \ang{3.7} \\
    EchoNet3D & \reviewbox{\ang{1.6}} & \reviewbox{\ang{1.6}} & \reviewbox{\ang{15.1}} & \reviewbox{0.27} & \reviewbox{0.26} & \reviewbox{0.17} & \reviewbox{\ang{2.3}} & \ang{0.3} & \reviewbox{0.39} & 0.05 & \reviewbox{\ang{4.6}} \\
    MITEA & \reviewbox{\ang{1.5}} & \ang{1.8} & \reviewbox{\ang{13.0}} & \reviewbox{0.25} & \reviewbox{0.30} & \reviewbox{0.14} & \reviewbox{\ang{3.5}} & \reviewbox{\ang{1.2}} & \reviewbox{0.64} & \reviewbox{0.19} & \reviewbox{\ang{4.7}} \\
    STACOM & \reviewbox{\ang{2.3}} & \ang{2.9} & \reviewbox{\ang{20.4}} & \reviewbox{0.40} & \reviewbox{0.51} & \reviewbox{0.23} & \ang{4.0} & \reviewbox{\ang{0.4}} & \reviewbox{0.75} & \reviewbox{0.06} & \reviewbox{\ang{11.4}} \\
    \bottomrule
    \end{tabular*}
\end{table*}

\hl{Both the heatmaps and the table report relative sweep error after centring each frame and axis.} \hl{Removing only the imposed starting rotation instead raised the EchoXFlow beam-axis error from 8.6° to 11.0°.}
\hl{Rotation error about} \reviewmath{x} \hl{grew with the applied angle, from 0.8° at 0--5° to 1.7° at 5--10° on EchoXFlow. Near the sweep limits, probe rotations about $x$ and $y$ were underestimated, consistent with uncertain estimates shrinking toward the centre of the training range. Bootstrap 95\% CIs over recordings for the beam-axis error were 7.4--10.0° (EchoXFlow), 9.8--13.0° (CETUS), 11.8--18.4° (EchoNet3D), 11.9--14.2° (MITEA), and 13.4--29.1° (STACOM). Beam-axis errors were highest on STACOM, driven largely by three of 15 subjects whose probe position appeared distant from the LV apex. Translation error was larger about} \reviewmath{x}\hl{ than about} \reviewmath{y}\hl{, consistent with tilts about} \reviewmath{x}\hl{ moving the sampling plane out of itself and tilts about} \reviewmath{y}\hl{ acting within it (see} \reviewbox{\Cref{fig:prob-form}} \hl{for the coordinate convention).}

\subsection{Annotated apical views in 2D and 3D}
\begin{figure*}[t]
    \centering
    \makebox[0.57\linewidth][c]{\includegraphics[width=0.223\linewidth]{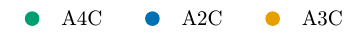}}\hfill\makebox[0.384\linewidth][c]{}\\[-15pt]
    \begin{minipage}[t]{0.28\linewidth}
        \centering
        \includegraphics[width=\linewidth]{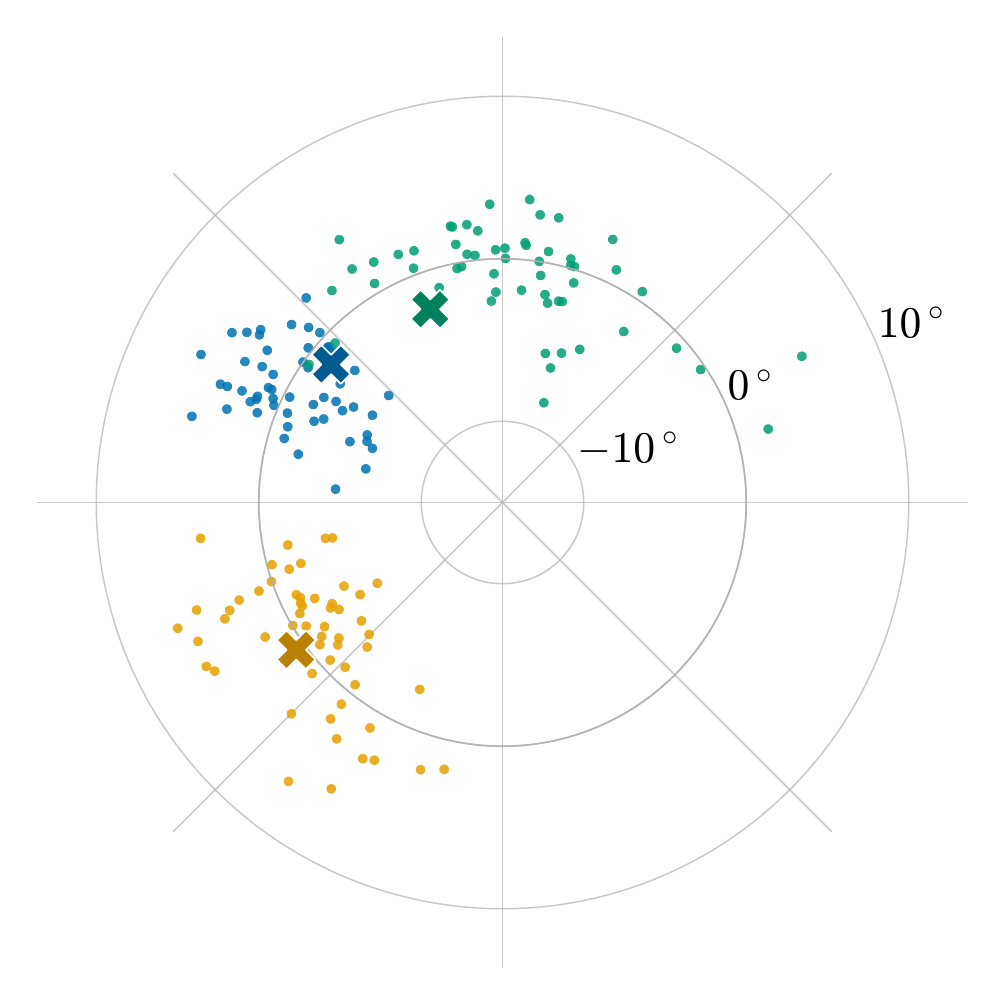}
    \end{minipage}
    \hfill
    \begin{minipage}[t]{0.28\linewidth}
        \centering
        \includegraphics[width=\linewidth]{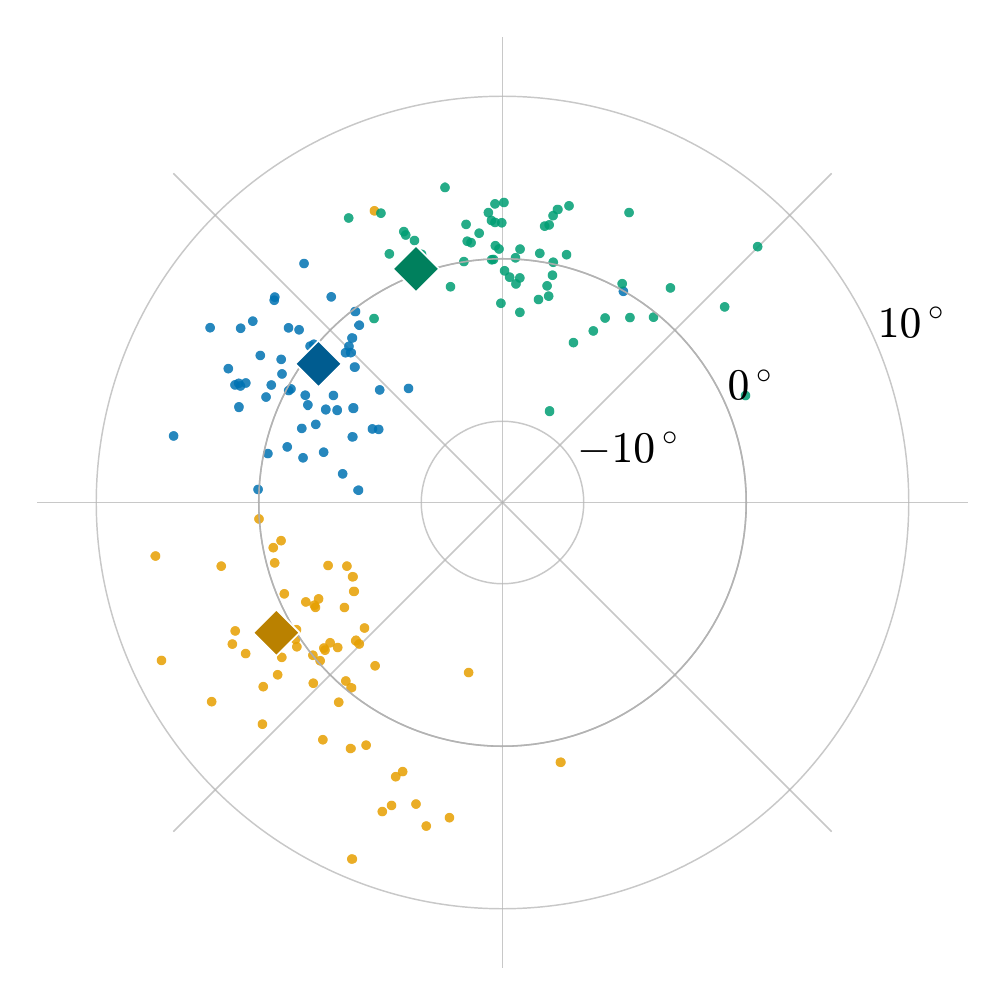}
    \end{minipage}
    \hfill
    \begin{minipage}[t]{0.395\linewidth}
        \centering
        \includegraphics[width=\linewidth]{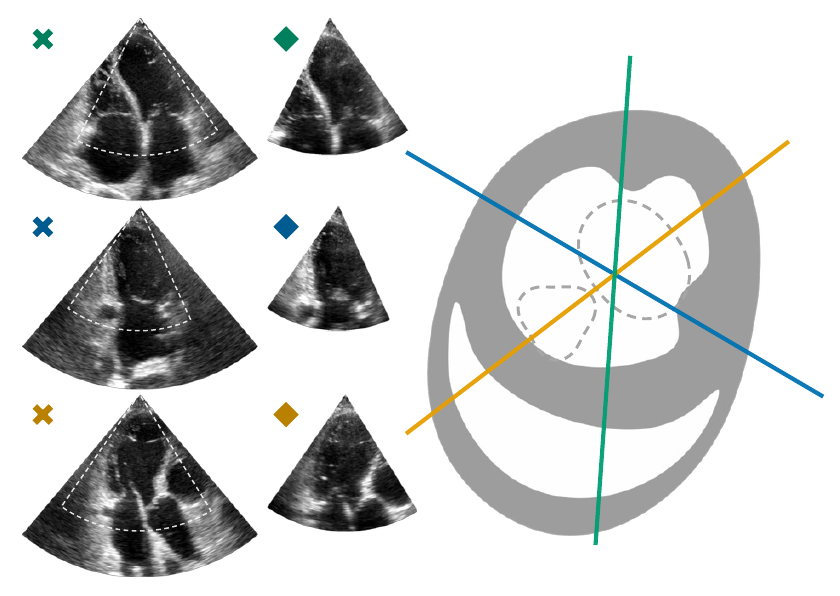}
    \end{minipage}
    \caption{Predicted probe poses for annotated apical views, comparing \hl{resampled native} 2D frames (left) against their paired annotated 3D reslices (middle) under the same angular convention and axis scale; colour denotes the apical view. Right: one exam's co-registered end-diastolic pairs (rows: A4C, A2C, A3C), each showing the \hl{native} 2D frame \hl{as acquired} beside its 3D reslice, and a stylised apical short-axis cross-section with their imaging planes.}
    \label{fig:exp1-polar-pose}
\end{figure*}

To visualise the cardiac structure captured by the learned representation, \Cref{fig:exp1-polar-pose} plots pose predictions for the annotated apical views of the evaluation pairs (\Cref{sec:coregistered-pairs}), comparing each \hl{native} 2D cine against its co-registered 3D reslice. \hl{Native images were resampled to match the depth, sector width, and centring of their reslices.} Each point summarises one cine by the $\mathrm{SO}(3)$-mean of its per-frame predictions, drawn at the predicted pose. Although the training loss constrains only pose differences, the three apical views form distinct clusters that are consistent across patients and across the two modalities. \hl{Native 2D probe-pose clusters were tighter than reslice clusters for A2C and A3C but wider for A4C, measured by angular distance to each view's mean pose in other individuals.} \hl{We also compared predicted probe rotations with the annotated plane orientations, after aligning their coordinate frames with one rotation fitted on other individuals. Mean disagreement for native versus resliced images was 17° vs.~15°, 18° vs.~16°, and 18° vs.~13° for A2C, A3C, and A4C, respectively. The annotated plane is itself an estimate from a separate acquisition.}

\subsection{Intra-patient guidance simulation}\label{sec:intra-patient-guidance}
Reproducible acquisition of standard 2D views within the same patient is essential to distinguish between true changes in cardiac function and variability due to inconsistent imaging planes~\cite{mitchellGuidelinesPerformingComprehensive2019,langRecommendationsCardiacChamber2015}, particularly in serial examinations such as cardiotoxicity monitoring and valvular disease follow-up~\cite{lyon2022ESCGuidelines2022,vahanian2021ESCEACTS2022}. A necessary step for such reproducibility is the ability to consistently recover a target view regardless of the initial probe pose. We therefore evaluate guided view recovery, where a controller attempts to guide a virtual probe from an initial random pose toward a specific apical view, using only model predictions.

To define the control objective, pose inference is first performed on a 2D ultrasound cine depicting the desired apical view. This cine is acquired separately but from the same patient, yielding the target pose prediction, \reviewmath{F_\Theta(x^{\dagger})}. During the simulation, a 2D ultrasound slice is sampled from the current probe pose at each time step, and the corresponding predicted pose, \reviewmath{F_\Theta(x_t)}, is inferred. The control objective is to minimise the predicted pose error,

\[
\bar{e}_{x^\dagger}(x_t) = d\left(\reviewmath{F_\Theta(x_t)}, \reviewmath{F_\Theta(x^{\dagger})}\right).
\]

\begin{figure*}[t]
    \centering
    \includegraphics[width=\textwidth]{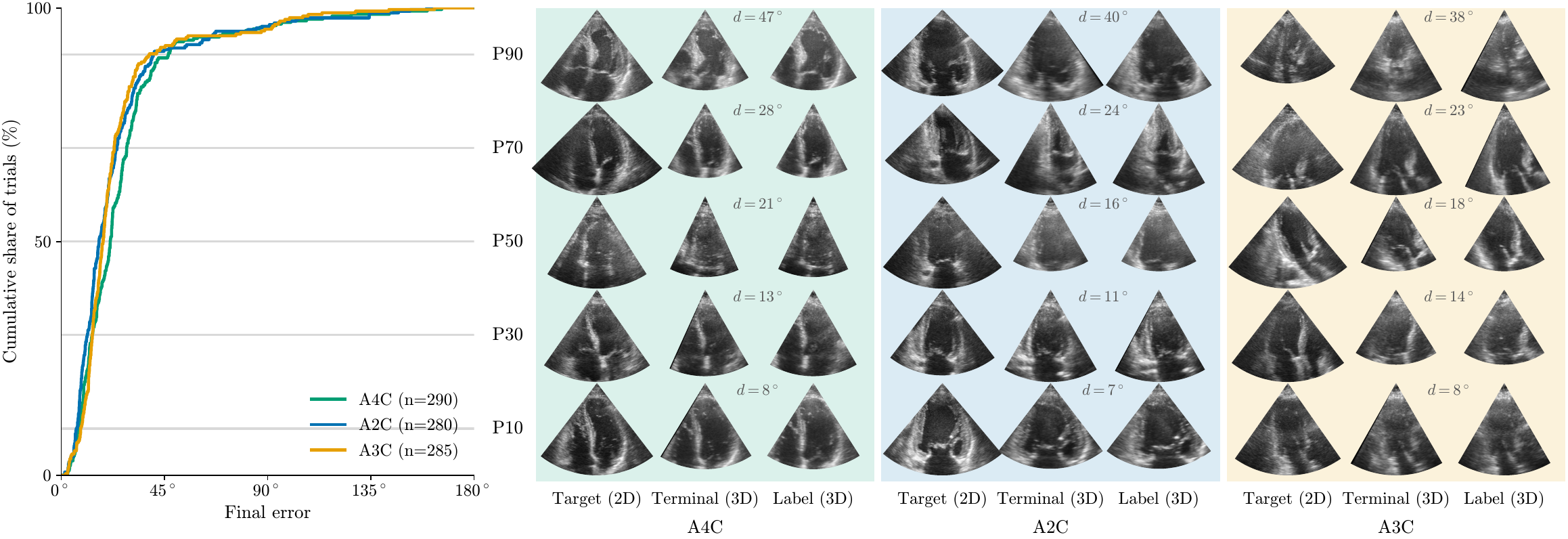}
    \caption{Rotation-only view recovery over the dynamic cardiac cycle, scored against the annotated probe pose. Left: empirical cumulative distribution of the final true error per target view. Right: example trials at the P10--P90 percentiles of final error for each view, showing a frame of the 2D target cine alongside the terminal and annotated label poses resliced at end-diastole; panels within a triplet share a physical scale, and each triplet is scaled to its row. The annotation above each triplet is the geodesic distance $d$ between the terminal and label poses.}
    \label{fig:experiment2a-ecdf}
\end{figure*}

\hl{Each update applies the predicted correction, capped at 3° per controlled rotation: probe rotation here, probe translation in} \reviewbox{\Cref{sec:world-tilt-guidance}}\hl{, and both in} \reviewbox{\Cref{sec:inter-patient-guidance}}\hl{. The model score is the predicted error, using the larger when both are controlled. A trial stops after 100 updates, or earlier if the score stalls (under 0.25° gain over ten updates) or the pose cycles (returns within 0.25° of one of the last 20 without improving). It has converged if the final score is below 5°, a pragmatic threshold.} In this setup, the probe position is held fixed near the apex, and only rotation is considered. Guidance to each target is simulated five times, each from a random starting pose\hl{: offsets from the annotated probe pose were drawn uniformly within} \reviewmath{\pm10^\circ}\hl{ about} \reviewmath{x}\hl{ and} \reviewmath{y}\hl{ and} \reviewmath{\pm180^\circ}\hl{ about the beam axis, and redrawn until the model score exceeded 5°. Each trial was run with the volume held at one frame (static) and with the volume advancing by one frame per update (dynamic).} Because the 2D cine is registered within the 3D ultrasound volume, \hl{angular disagreement with the registered target plane can be computed directly} and used for quantitative evaluation. Results are summarised in \Cref{fig:experiment2a-ecdf}, also showing examples of terminal poses and the labelled poses at selected percentiles of \hl{the final error against the annotated pose}. \hl{In many high-error cases, the terminal and labelled poses still show qualitatively similar-looking anatomy.}

\subsection{\hl{Probe-translation} guidance simulation}\label{sec:world-tilt-guidance}
We repeat the guidance simulation of \Cref{sec:intra-patient-guidance}, but recover a known \hl{probe translation} rather than a target view: $\Rp$ is held fixed, and the controller corrects \hl{the translation} $\Rv$ (\Cref{sec:representation-of-rotations}) back toward the annotated pose. The imposed error is known by construction: each start tilts about a random axis perpendicular to the probe axis, with magnitude drawn uniformly between \ang{5} and \ang{15}. The controller stops on its own error estimate but is scored against the known tilt.

Results are summarised in \Cref{tab:experiment2b-controller}. Convergence is defined by model-scored error under \ang{5}. This occurs in \reviewbox{94.8\%}, \reviewbox{99.3\%}, and \reviewbox{95.8\%} for A4C, A2C, and A3C \hl{(95\% CIs over individuals: 91.0--97.9, 97.8--100, and 91.3--99.0)}. Converged trials end below \reviewbox{\ang{1}} by the model's own estimate, against true residuals of \reviewbox{3}--\reviewbox{\ang{4}}.

\begin{table}[t]
    \caption{\hl{Probe-translation} recovery by target view, over the dynamic cardiac cycle, scored against the known imposed tilt. A trial converges if it rests below $5^\circ$ model-scored error. Error columns are means over converged trials only.}
    \label{tab:experiment2b-controller}
    \centering
    \footnotesize
    \setlength{\tabcolsep}{3pt}
    \begin{tabular*}{\linewidth}{@{\extracolsep{\fill}}lrcrrr@{}}
    \toprule
     & & & Convergence & \multicolumn{2}{c}{Error} \\
    \cmidrule(lr){5-6}
    Target & $n$ & Initial & (\%) & Model & True \\
    \midrule
        A4C & 290 & Unif(\ang{5},\ang{15}) & \reviewbox{94.8} & \ang{0.8} & \reviewbox{\ang{4.4}} \\
    A2C & 280 & Unif(\ang{5},\ang{15}) & \reviewbox{99.3} & \reviewbox{\ang{0.7}} & \reviewbox{\ang{2.6}} \\
    A3C & 285 & Unif(\ang{5},\ang{15}) & \reviewbox{95.8} & \ang{0.7} & \reviewbox{\ang{3.0}} \\
    \bottomrule

    \end{tabular*}
\end{table}

\hl{Of 855 dynamic trials, 90 started below the 5° threshold. At cutoffs of 2.5°, 5°, 7.5°, and 10°, 772, 826, 848, and 854 trials converged. Although the imposed tilt was known, the true error of converged trials exceeded the model score by 3.6° (95\% CI 2.7--4.6) for A4C, 2.0° (1.6--2.5) for A2C, and 2.3° (1.7--3.0) for A3C.}

\subsection{Inter-patient guidance simulation}\label{sec:inter-patient-guidance}
\begin{figure*}[t]
    \centering
    \includegraphics[width=\textwidth]{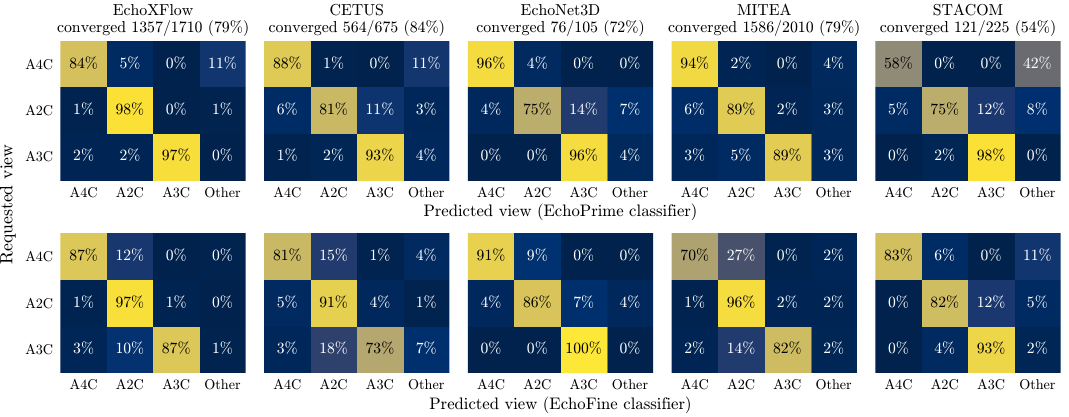}
    \caption{Terminal-frame view classification as row-normalised confusion matrices, for EchoPrime (top) and EchoFine (bottom). Columns are EchoXFlow (in-domain reference) followed by the four external test sets; panel titles report model-scored convergence as converged/attempted trials per dataset. Within each panel, rows are the requested view and columns the classifier's collapsed prediction among converged trials. External panels show \hl{joint guidance of probe rotation and translation}; the EchoXFlow reference column pools the static and dynamic conditions of the rotation-only trials of \Cref{sec:intra-patient-guidance}.}
    \label{fig:exp3-confusion}
\end{figure*}

\hl{For each apical view, we built a consensus target from the matched EchoXFlow test examples and used it to guide probe rotation and translation in the four external datasets.} \hl{External trials started at zero translation offset. Probe offsets from the native volume orientation were drawn uniformly within} \reviewmath{\pm10^\circ}\hl{ about} \reviewmath{x}\hl{ and} \reviewmath{y}\hl{ and} \reviewmath{\pm180^\circ}\hl{ about the beam axis.} \hl{As a reference, the EchoXFlow column of the confusion matrix shows rotation-only guidance within the same patient, pooling 1,710 static and dynamic trials.} Results are summarised in \Cref{fig:exp3-confusion}. Among converged trials, both EchoPrime and EchoFine classify the terminal frame as the requested view in the majority of cases. The two classifiers showed substantial agreement on the external datasets (Cohen's $\kappa$ of \reviewbox{0.73}). \hl{STACOM has the lowest convergence rate in this simulation.}

\hl{Of 3,015 external attempts, 2,347 converged. Across all attempts, EchoPrime and EchoFine classified the terminal frame as the target in 77\% and 72\%, respectively. In 299 attempts (10\%), both agreed on another class, which was an apical view in 199 and Other in 100. The two classifiers disagreed with each other in 754 attempts (25\%).}

\subsection{\hl{Comparison with supervised baseline}}\label{sec:baseline}
\hl{We compare with the networks of Pasdeloup et al.}~\reviewbox{\cite{pasdeloupRealTimeEchocardiographyGuidance2023}}\hl{, trained on expert annotations to predict probe rotation and view-specific tilt. Both methods moved a virtual probe with these two motions from the same 4,725 starts, in steps of at most 3° for up to 100 updates. Applying their operator cues to a simulated probe is our adaptation.} \reviewbox{\Cref{tab:baseline}} \hl{shows how often the terminal frame was classified as the target view. Echo-POSED was significantly better on EchoXFlow and MITEA, and Pasdeloup et al. on STACOM with EchoFine. Differences on CETUS and EchoNet3D were not significant.}

\begin{table}[t]
    \caption{\hl{Target-view agreement (\%) of terminal frames.} \reviewmath{p}\hl{: Holm-adjusted, paired permutation test over individuals. Bold: significantly better method,} \reviewmath{p<0.05}\hl{.} \reviewmath{^\dagger}\hl{Used in developing the model of Pasdeloup et al.}}
    \label{tab:baseline}
    \centering
    \scriptsize
    \setlength{\tabcolsep}{1pt}
    \begin{tabular*}{\linewidth}{@{\extracolsep{\fill}}lccr@{\extracolsep{0pt}}l@{\extracolsep{\fill}}ccr@{\extracolsep{0pt}}l@{}}
    \toprule
     & \multicolumn{4}{c}{\hl{EchoPrime}} & \multicolumn{4}{c}{\hl{EchoFine}} \\
    \cmidrule(lr){2-5} \cmidrule(l){6-9}
    \hl{Dataset} & \hl{Pasdeloup et al.} & \hl{Ours} & \multicolumn{2}{c}{\reviewmath{p}} & \hl{Pasdeloup et al.} & \hl{Ours} & \multicolumn{2}{c}{\reviewmath{p}} \\
    \midrule
    \hl{EchoXFlow} & \hl{70} & \textbf{\hl{82}} & \hl{0} & \hl{.001} & \hl{70} & \textbf{\hl{78}} & \hl{0} & \hl{.007} \\
    \hl{CETUS}\reviewmath{^\dagger} & \hl{85} & \hl{83} & \hl{1} & \hl{.00} & \hl{76} & \hl{72} & \hl{0} & \hl{.46} \\
    \hl{EchoNet3D} & \hl{77} & \hl{72} & \hl{1} & \hl{.00} & \hl{82} & \hl{72} & \hl{0} & \hl{.75} \\
    \hl{MITEA} & \hl{63} & \textbf{\hl{80}} & \reviewmath{<0} & \hl{.001} & \hl{62} & \textbf{\hl{67}} & \hl{0} & \hl{.02} \\
    \hl{STACOM} & \hl{71} & \hl{61} & \hl{0} & \hl{.19} & \textbf{\hl{76}} & \hl{64} & \hl{0} & \hl{.047} \\
    \bottomrule
    \end{tabular*}
\end{table}

\subsection{\hl{Ablations and control analyses}}\label{sec:ablations}
\hl{To isolate training choices, we retrained the four-member ensemble from ImageNet initialisation with one change per run: drawing all slices of a batch from the same cardiac phase; removing the corner-sector mask; or removing sector-width reduction.} \reviewbox{\Cref{tab:training-ablation}}\hl{ reports errors on identical mixed perturbations of the held-out recordings. Removing cross-phase sampling had the largest effect on probe rotation, and increased both errors. Removing sector and corner augmentations did not uniformly degrade performance, as expected, because their purpose is to prohibit the model from ``cheating'' and using the sector geometry rather than universally improving performance. On the resampled 2D--3D pairs, removing the corner mask increased mean translation disagreement from 1.1° to 3.2° (paired difference 2.1°, 95\% CI 1.6 to 2.5). This is consistent with the unmasked model relying on sector shape.}

\begin{table}[t]
    \caption{\hl{Training ablations. Mean geodesic error ($^\circ$) on 7,296 mixed perturbations, with 95\% CIs of paired differences.}}
    \label{tab:training-ablation}
    \centering
    \scriptsize
    \setlength{\tabcolsep}{2pt}
    \begin{tabular*}{\linewidth}{@{\extracolsep{\fill}}llr@{\extracolsep{0pt}\,}l@{\extracolsep{\fill}}lr@{\extracolsep{0pt}\,}l@{}}
    \toprule
     & \multicolumn{3}{c}{\hl{Probe translation} \reviewmath{\Rv}} & \multicolumn{3}{c}{\hl{Probe rotation} \reviewmath{\Rp}} \\
    \cmidrule(lr){2-4} \cmidrule(l){5-7}
    \hl{Training run} & \hl{Error} & \multicolumn{2}{c}{\hl{Difference (95\% CI)}} & \hl{Error} & \multicolumn{2}{c}{\hl{Difference (95\% CI)}} \\
    \midrule
    \hl{Echo-POSED (reference)} & \hl{6.22} & & & \hl{11.60} & & \\
    \hl{No cross-phase sampling} & \hl{6.45} & \reviewmath{[0.14,} & \reviewmath{0.32]} & \hl{13.47} & \reviewmath{[1.23,} & \reviewmath{2.52]} \\
    \hl{No corner mask} & \hl{5.81} & \reviewmath{[-0.51,} & \reviewmath{-0.32]} & \hl{12.47} & \reviewmath{[0.45,} & \reviewmath{1.30]} \\
    \hl{No sector-width reduction} & \hl{6.01} & \reviewmath{[-0.29,} & \reviewmath{-0.14]} & \hl{11.38} & \reviewmath{[-0.70,} & \reviewmath{0.29]} \\
    \bottomrule
    \end{tabular*}
\end{table}

\hl{We also offset probe rotation and translation simultaneously, in a pilot with 108 starts from 12 held-out recordings. Controlling both reduced mean translation error from 10° to 4° and mean rotation error from 36° to 13°. Controlling only one left the other uncorrected, but reached a lower error in the controlled one: 11° for rotation and 2° for translation.}

\hl{The shape of a resliced sector varies with pose, so the model could infer pose from this shape rather than from anatomy. To test this, we kept only the sector mask, or restricted anatomy to a field of view shared by all poses. Under joint perturbation, translation and rotation errors were 6° and 12° with the original images, 10° and 76° with the mask alone, and 10° and 13° with the shared field of view. Probe rotation was thus poorly recovered from the mask alone; the test does not exclude a contribution of sector shape to translation estimates.}

\hl{Finally, we examined how much ensembling and member alignment contribute to accuracy. Averaged over all member subsets, rotation error was 12.9°, 12.2°, 11.8°, and 11.6° for one to four members. Without member alignment, translation and rotation errors rose from 6.2° and 11.6° to 7.4° and 13.0°.}

\reviewbox{\Cref{tab:rotation-head-ablation}} \hl{shows that 9D SVD and 6D Gram--Schmidt had nearly identical test errors, both far lower than those of the discontinuous representations} (\reviewbox{\Cref{tab:so3_representations}}).

\begin{table}[t]
    \caption{\hl{Rotation representation ablation. Summed test error of relative probe rotation and translation at lowest validation error.}}
    \label{tab:rotation-head-ablation}
    \centering
    \footnotesize
    \begin{tabular*}{\linewidth}{@{\extracolsep{\fill}}lrr@{}}
    \toprule
    \hl{Representation} & \hl{Error ($^\circ$)} & \hl{95\% CI} \\
    \midrule
    \hl{Euler}                   & \hl{31.8} & \hl{28.6--35.2} \\
    \hl{Exponential coordinates} & \hl{31.1} & \hl{28.0--34.6} \\
    \hl{Unit quaternions}        & \hl{32.7} & \hl{29.6--36.3} \\
    \hl{Axis--angle}             & \hl{31.1} & \hl{28.0--34.5} \\
    \hl{6D + Gram--Schmidt} & \hl{19.8} & \hl{16.8--23.4} \\
    \textbf{\hl{9D + singular value decomposition}} & \textbf{\hl{19.2}} & \textbf{\hl{16.3--22.6}} \\
    \bottomrule
    \end{tabular*}
\end{table}

\section{Discussion}
Our experiments support two key capabilities. First, Echo-POSED establishes a consistent pose representation that generalises across patients and acquisition hardware, despite being trained in a completely self-supervised manner, as demonstrated by angular calibration under controlled virtual perturbations (\Cref{sec:ang-regression}). This representation also appears to be stable across patients, as illustrated in~\Cref{fig:exp1-polar-pose}.
Second, the representation supports closed-loop guidance, illustrated in \Cref{fig:experiment2a-ecdf} where the target was a 2D cine from the same patient, and in \Cref{fig:exp3-confusion} where the target was selected as a consensus view from other patients.

\hl{We next ask whether ensemble disagreement indicates how reliable an estimate is. Datasets with higher mean disagreement generally had larger beam-axis sweep errors} (\reviewbox{\Cref{tab:scatter-summarised}})\hl{; only EchoNet3D and MITEA, with similar disagreement, exchange places. Within the sweeps, error also rose on average with disagreement} (\reviewbox{\Cref{fig:experiment1-uncertainty-calibration}})\hl{. Disagreement is thus a useful relative indicator of reliability, but not a calibrated measure of absolute error, since members can agree and still be wrong.}

\hl{This work has several limitations. First, no evaluation used a tracked physical probe: guidance was simulated by reslicing 3D volumes, and the 2D--3D comparison relies on a plane estimated from a separate acquisition, without repeat annotations to measure reader variability. Second, the model score should be interpreted with caution. It cannot reveal errors in the predictions it is computed from, ensemble disagreement mainly reflects epistemic uncertainty, and in probe-translation recovery, where the imposed motion provides an independent reference, the controller underestimated the residual error. The score should therefore not be the sole acceptance criterion. Third, the classifier endpoints of} \reviewbox{\Cref{sec:inter-patient-guidance}} \hl{are imperfect: they measure agreement with the requested view label, not diagnostic quality. The same classifiers screened the EchoXFlow pairs, so agreement there is likely optimistic (the external datasets were not screened), whereas the two classifiers disagreed with each other on 25\% of external attempts, so their labels on resliced frames are themselves uncertain. Fourth, the model is specific to the apical window.} \hl{It assumes a pivot inside the volume, translation expressed as a small tilt, beam-axis rotation assigned to the probe, and sampling around the left ventricular long axis, none of which need hold for other windows. Translation tilts beyond the 10° used in training were tested only as starting offsets of up to 15° (\mbox{\Cref{sec:world-tilt-guidance}}), and we did not convert tilts into distances on the chest.} \hl{Probe translation can change the visible sector boundary in a reslice, potentially giving the model a cue to displacement relative to the original acquisition.} More broadly, Echo-POSED is a vision backbone rather than a complete guidance system; it does not model acoustic accessibility (ribs, lungs, window reachability), and a deployable system must build such components on top.

Future work includes extending training beyond the apical window (e.g., parasternal, subcostal, suprasternal) and improving the stability of human-in-the-loop guidance via uncertainty-aware filtering or sequence models. Another promising direction is to incorporate large-scale, publicly available 2D echocardiography data to improve stability around canonical standard views and to validate (and, if needed, fine-tune) recommendations against clinical acquisition guidelines. Beyond real-time assistance, a model that outputs both a guidance signal and an uncertainty estimate could support retrospective quality screening, training of novice sonographers, and integration into semi-autonomous or robotic ultrasound systems. Because the pretraining objective encourages reliance on stable anatomical cues, the learned representation may also transfer to downstream tasks such as landmark detection, view classification, and quality assessment.

\section{Conclusion}
Echo-POSED shows that geometric self-distillation on routinely acquired 3D echocardiography videos can be used to learn a globally aligned pose representation and corresponding probe-adjustment recommendations without any manual annotation or paired probe trajectories. By enforcing equivariance to virtual probe perturbations and invariance to cardiac phase, the model isolates pose-relevant anatomical variations from appearance changes driven by cardiac motion and other irrelevant information.

Finally, to enable reproducible research in this area, we release the Echo-POSED model together with our manually labelled, co-registered 2D and 3D poses from the EchoXFlow dataset as a public benchmark. These resources provide a common basis for reproducible evaluation and comparison of pose-estimation methods in echocardiography.

\section*{\hl{Code and Data Availability}}
\hl{Model code and weights are released as open-source, subject to the licences in the respective repositories.}\\
\url{https://github.com/Ahus-AIM/Echo-POSED}\\
\hl{Co-registered 2D--3D view pairs:}\\
\url{https://huggingface.co/datasets/Ahus-AIM/Echo-POSED-2D-3D-registration}
\raggedbottom
\bibliographystyle{IEEEtran}
\bibliography{IEEEtranBSTCTL,zotero_references,manual_references}

\begin{thebibliography}{10}
\providecommand{\url}[1]{#1}
\csname url@samestyle\endcsname
\providecommand{\newblock}{\relax}
\providecommand{\bibinfo}[2]{#2}
\providecommand{\BIBentrySTDinterwordspacing}{\spaceskip=0pt\relax}
\providecommand{\BIBentryALTinterwordstretchfactor}{4}
\providecommand{\BIBentryALTinterwordspacing}{\spaceskip=\fontdimen2\font plus
\BIBentryALTinterwordstretchfactor\fontdimen3\font minus
  \fontdimen4\font\relax}
\providecommand{\BIBforeignlanguage}[2]{{%
\expandafter\ifx\csname l@#1\endcsname\relax
\typeout{** WARNING: IEEEtran.bst: No hyphenation pattern has been}%
\typeout{** loaded for the language `#1'. Using the pattern for}%
\typeout{** the default language instead.}%
\else
\language=\csname l@#1\endcsname
\fi
#2}}
\providecommand{\BIBdecl}{\relax}
\BIBdecl

\bibitem{langRecommendationsCardiacChamber2015}
R.~M. Lang \emph{et~al.}, ``Recommendations for {Cardiac} {Chamber}
  {Quantification} by {Echocardiography} in {Adults}: {An} {Update} from the
  {American} {Society} of {Echocardiography} and the {European} {Association}
  of {Cardiovascular} {Imaging},'' \emph{Journal of the American Society of
  Echocardiography}, vol.~28, no.~1, pp. 1--39.e14, Jan. 2015.

\bibitem{hagendorffValidReproducibleQuantitative2023}
A.~Hagendorff, J.~Kandels, M.~Metze, B.~Tayal, and S.~Stöbe,
  ``\BIBforeignlanguage{en}{Valid and {Reproducible} {Quantitative}
  {Assessment} of {Cardiac} {Volumes} by {Echocardiography} in {Patients} with
  {Valvular} {Heart} {Diseases}—{Possible} or {Wishful} {Thinking}?}''
  \emph{\BIBforeignlanguage{en}{Diagnostics}}, vol.~13, no.~7, p. 1359, Apr.
  2023.

\bibitem{mitchellGuidelinesPerformingComprehensive2019}
C.~Mitchell \emph{et~al.}, ``\BIBforeignlanguage{en}{Guidelines for
  {Performing} a {Comprehensive} {Transthoracic} {Echocardiographic}
  {Examination} in {Adults}: {Recommendations} from the {American} {Society} of
  {Echocardiography}},'' \emph{\BIBforeignlanguage{en}{Journal of the American
  Society of Echocardiography}}, vol.~32, no.~1, pp. 1--64, Jan. 2019.

\bibitem{kirkpatrickRecommendationsEchocardiographyLaboratories2020}
J.~N. Kirkpatrick \emph{et~al.}, ``\BIBforeignlanguage{eng}{Recommendations for
  {Echocardiography} {Laboratories} {Participating} in {Cardiac} {Point} of
  {Care} {Cardiac} {Ultrasound} ({POCUS}) and {Critical} {Care}
  {Echocardiography} {Training}: {Report} from the {American} {Society} of
  {Echocardiography}},'' \emph{\BIBforeignlanguage{eng}{Journal of the American
  Society of Echocardiography}}, vol.~33, no.~4, pp. 409--422.e4, Apr. 2020.

\bibitem{biMachineLearningRobotic2024}
Y.~Bi, Z.~Jiang, F.~Duelmer, D.~Huang, and N.~Navab,
  ``\BIBforeignlanguage{en}{Machine {Learning} in {Robotic} {Ultrasound}
  {Imaging}: {Challenges} and {Perspectives}},''
  \emph{\BIBforeignlanguage{en}{Annual Review of Control, Robotics, and
  Autonomous Systems}}, vol.~7, no.~1, pp. 335--357, Jul. 2024.

\bibitem{yue_echoworld_2025}
Y.~Yue, Y.~Wang, H.~Jiang, P.~Liu, S.~Song, and G.~Huang, ``{EchoWorld}:
  {Learning} {Motion}-{Aware} {World} {Models} for {Echocardiography} {Probe}
  {Guidance},'' in \emph{2025 {IEEE}/{CVF} {Conference} on {Computer} {Vision}
  and {Pattern} {Recognition} ({CVPR})}, Jun. 2025, pp. 25\,993--26\,003.

\bibitem{pasdeloupRealTimeEchocardiographyGuidance2023}
D.~Pasdeloup \emph{et~al.}, ``\BIBforeignlanguage{eng}{Real-{Time}
  {Echocardiography} {Guidance} for {Optimized} {Apical} {Standard} {Views}},''
  \emph{\BIBforeignlanguage{eng}{Ultrasound in Medicine \& Biology}}, vol.~49,
  no.~1, pp. 333--346, Jan. 2023.

\bibitem{ferrazAssistedProbeGuidance2023}
S.~Ferraz, M.~Coimbra, and J.~Pedrosa, ``Assisted probe guidance in cardiac
  ultrasound: {A} review,'' \emph{Frontiers in Cardiovascular Medicine},
  vol.~10, p. 1056055, Feb. 2023.

\bibitem{gopalakrishnanIntraoperative2D3D2024}
V.~Gopalakrishnan, N.~Dey, and P.~Golland, ``Intraoperative {2D}/{3D} {Image}
  {Registration} via {Differentiable} {X}-ray {Rendering},'' in \emph{2024
  {IEEE}/{CVF} {Conference} on {Computer} {Vision} and {Pattern} {Recognition}
  ({CVPR})}, Jun. 2024, pp. 11\,662--11\,672.

\bibitem{huhAIDrivenViewGuidance2025}
J.~Huh, P.~Klein, G.~Funka-Lea, P.~Sharma, A.~Kapoor, and Y.-H. Kim,
  ``\BIBforeignlanguage{eng}{{AI}-{Driven} {View} {Guidance} {System} in
  {Intra}-{Cardiac} {Echocardiography} {Imaging}},''
  \emph{\BIBforeignlanguage{eng}{IEEE Transactions on Biomedical Engineering}},
  vol.~72, no.~7, pp. 2072--2084, Jul. 2025.

\bibitem{madaniFastAccurateView2018}
A.~Madani, R.~Arnaout, M.~Mofrad, and R.~Arnaout,
  ``\BIBforeignlanguage{en}{Fast and accurate view classification of
  echocardiograms using deep learning},'' \emph{\BIBforeignlanguage{en}{npj
  Digital Medicine}}, vol.~1, no.~1, p.~6, Mar. 2018.

\bibitem{ostvikRealTimeStandardView2019}
A.~Østvik, E.~Smistad, S.~A. Aase, B.~O. Haugen, and L.~Lovstakken,
  ``\BIBforeignlanguage{eng}{Real-{Time} {Standard} {View} {Classification} in
  {Transthoracic} {Echocardiography} {Using} {Convolutional} {Neural}
  {Networks}},'' \emph{\BIBforeignlanguage{eng}{Ultrasound in Medicine \&
  Biology}}, vol.~45, no.~2, pp. 374--384, Feb. 2019.

\bibitem{kimAutomatedDetectionApical2023}
W.-J.~C. Kim \emph{et~al.}, ``Automated {Detection} of {Apical}
  {Foreshortening} in {Echocardiography} {Using} {Statistical} {Shape}
  {Modelling},'' \emph{Ultrasound in Medicine \& Biology}, vol.~49, no.~9, pp.
  1996--2005, Sep. 2023.

\bibitem{smistadRealTimeAutomaticEjection2020}
E.~Smistad \emph{et~al.}, ``\BIBforeignlanguage{eng}{Real-{Time} {Automatic}
  {Ejection} {Fraction} and {Foreshortening} {Detection} {Using} {Deep}
  {Learning}},'' \emph{\BIBforeignlanguage{eng}{IEEE Transactions on
  Ultrasonics, Ferroelectrics, and Frequency Control}}, vol.~67, no.~12, pp.
  2595--2604, Dec. 2020.

\bibitem{narangUtilityDeepLearningAlgorithm2021}
A.~Narang \emph{et~al.}, ``\BIBforeignlanguage{en}{Utility of a
  {Deep}-{Learning} {Algorithm} to {Guide} {Novices} to {Acquire}
  {Echocardiograms} for {Limited} {Diagnostic} {Use}},''
  \emph{\BIBforeignlanguage{en}{JAMA Cardiology}}, vol.~6, no.~6, pp. 624--632,
  Jun. 2021.

\bibitem{baoRealWorldVisualNavigation2024}
M.~Bao \emph{et~al.}, ``\BIBforeignlanguage{en}{Real-{World} {Visual}
  {Navigation} for {Cardiac} {Ultrasound} {View} {Planning}},'' in
  \emph{\BIBforeignlanguage{en}{Medical {Image} {Computing} and {Computer}
  {Assisted} {Intervention} – {MICCAI} 2024}}, M.~G. Linguraru \emph{et~al.},
  Eds.\hskip 1em plus 0.5em minus 0.4em\relax Cham: Springer Nature
  Switzerland, 2024, pp. 317--326.

\bibitem{jiangStructureawareWorldModel2025}
H.~Jiang \emph{et~al.}, ``\BIBforeignlanguage{en}{Structure-aware {World}
  {Model} for {Probe} {Guidance} via {Large}-scale {Self}-supervised
  {Pre}-train},'' in \emph{\BIBforeignlanguage{en}{Simplifying {Medical}
  {Ultrasound}}}, A.~Gomez, B.~Khanal, A.~King, and A.~Namburete, Eds.\hskip
  1em plus 0.5em minus 0.4em\relax Cham: Springer Nature Switzerland, 2025, pp.
  58--67.

\bibitem{saboRealtimeGuidingDeep2023}
S.~Sabo \emph{et~al.}, ``\BIBforeignlanguage{en}{Real-time guiding by deep
  learning during echocardiography to reduce left ventricular foreshortening
  and measurement variability},'' \emph{\BIBforeignlanguage{en}{European Heart
  Journal - Imaging Methods and Practice}}, vol.~1, no.~1, p. qyad012, May
  2023.

\bibitem{saboRealtimeGuidanceDeep2023}
------, ``Real-time guidance by deep learning of experienced operators to
  improve the standardization of echocardiographic acquisitions,''
  \emph{European Heart Journal - Imaging Methods and Practice}, vol.~1, no.~2,
  p. qyad040, Nov. 2023.

\bibitem{liStandardPlaneLocalisation2018}
Y.~Li \emph{et~al.}, ``\BIBforeignlanguage{en}{Standard {Plane} {Localisation}
  in {3D} {Fetal} {Ultrasound} {Using} {Network} with {Geometric} and {Image}
  {Loss}},'' \emph{\BIBforeignlanguage{en}{Medical Imaging with Deep
  Learning}}, Apr. 2018.

\bibitem{liStandardPlaneDetection2018}
------, ``\BIBforeignlanguage{en}{Standard {Plane} {Detection} in {3D} {Fetal}
  {Ultrasound} {Using} an {Iterative} {Transformation} {Network}},'' in
  \emph{\BIBforeignlanguage{en}{Medical {Image} {Computing} and {Computer}
  {Assisted} {Intervention} – {MICCAI} 2018}}, A.~F. Frangi, J.~A. Schnabel,
  C.~Davatzikos, C.~Alberola-López, and G.~Fichtinger, Eds.\hskip 1em plus
  0.5em minus 0.4em\relax Cham: Springer International Publishing, 2018, pp.
  392--400.

\bibitem{menScanAheadSimplifyingStandard2025}
Q.~Men, H.~Zhao, L.~Drukker, A.~T. Papageorghiou, and J.~A. Noble,
  ``{ScanAhead}: {Simplifying} standard plane acquisition of fetal head
  ultrasound,'' \emph{Medical Image Analysis}, vol. 104, p. 103614, Aug. 2025.

\bibitem{linguraru_pose-guidenet_2024}
Q.~Men, X.~Guo, A.~T. Papageorghiou, and J.~A. Noble,
  ``\BIBforeignlanguage{en}{Pose-{GuideNet}: {Automatic} {Scanning} {Guidance}
  for {Fetal} {Head} {Ultrasound} from {Pose} {Estimation}},'' in
  \emph{\BIBforeignlanguage{en}{Medical {Image} {Computing} and {Computer}
  {Assisted} {Intervention} – {MICCAI} 2024}}, ser. Lecture {Notes} in
  {Computer} {Science}, M.~G. Linguraru \emph{et~al.}, Eds.\hskip 1em plus
  0.5em minus 0.4em\relax Cham: Springer Nature Switzerland, 2024, vol. 15004,
  pp. 700--710.

\bibitem{droste_automatic_2020}
R.~Droste, L.~Drukker, A.~T. Papageorghiou, and J.~A. Noble,
  ``\BIBforeignlanguage{en}{Automatic {Probe} {Movement} {Guidance} for
  {Freehand} {Obstetric} {Ultrasound}},'' in
  \emph{\BIBforeignlanguage{en}{Medical {Image} {Computing} and {Computer}
  {Assisted} {Intervention} – {MICCAI} 2020}}, A.~L. Martel \emph{et~al.},
  Eds.\hskip 1em plus 0.5em minus 0.4em\relax Cham: Springer International
  Publishing, 2020, pp. 583--592.

\bibitem{yeungLearningMap2D2021}
P.-H. Yeung, M.~Aliasi, A.~T. Papageorghiou, M.~Haak, W.~Xie, and A.~I.~L.
  Namburete, ``Learning to map {2D} ultrasound images into {3D} space with
  minimal human annotation,'' \emph{Medical Image Analysis}, vol.~70, p.
  101998, May 2021.

\bibitem{midtvedtSingleshotSelfsupervisedObject2022}
B.~Midtvedt \emph{et~al.}, ``\BIBforeignlanguage{en}{Single-shot
  self-supervised object detection in microscopy},''
  \emph{\BIBforeignlanguage{en}{Nature Communications}}, vol.~13, no.~1, p.
  7492, Dec. 2022.

\bibitem{nakataniLeftVentricularRotation2011}
S.~Nakatani, ``\BIBforeignlanguage{en}{Left {Ventricular} {Rotation} and
  {Twist}: {Why} {Should} {We} {Learn}?}''
  \emph{\BIBforeignlanguage{en}{Journal of Cardiovascular Ultrasound}},
  vol.~19, no.~1, p.~1, 2011.

\bibitem{zhou_continuity_2019}
Y.~Zhou, C.~Barnes, J.~Lu, J.~Yang, and H.~Li, ``On the {Continuity} of
  {Rotation} {Representations} in {Neural} {Networks},'' in \emph{2019
  {IEEE}/{CVF} {Conference} on {Computer} {Vision} and {Pattern} {Recognition}
  ({CVPR})}, Jun. 2019, pp. 5738--5746.

\bibitem{levinson_analysis_2020}
J.~Levinson \emph{et~al.}, ``An {Analysis} of {SVD} for {Deep} {Rotation}
  {Estimation},'' in \emph{Advances in {Neural} {Information} {Processing}
  {Systems}}, vol.~33.\hskip 1em plus 0.5em minus 0.4em\relax Curran
  Associates, Inc., 2020, pp. 22\,554--22\,565.

\bibitem{geist_learning_2024}
A.~R. Geist, J.~Frey, M.~Zhobro, A.~Levina, and G.~Martius,
  ``\BIBforeignlanguage{en}{Learning with {3D} rotations, a hitchhiker’s
  guide to {SO}(3)},'' in \emph{\BIBforeignlanguage{en}{Proceedings of the 41st
  {International} {Conference} on {Machine} {Learning}}}.\hskip 1em plus 0.5em
  minus 0.4em\relax PMLR, Jul. 2024, pp. 15\,331--15\,350.

\bibitem{stenhedeEchoXFlowBeamspaceEchocardiography2026}
E.~Stenhede, J.~Sulkowska, E.~B. Orstad, H.~Schirmer, and A.~Ranjbar,
  ``{EchoXFlow}: {A} {Beamspace} {Echocardiography} {Dataset} for {Cardiac}
  {Motion}, {Flow}, and {Function},'' May 2026, arXiv:2605.05447 [cs.CV].

\bibitem{bernard_standardized_2016}
O.~Bernard \emph{et~al.}, ``Standardized {Evaluation} {System} for {Left}
  {Ventricular} {Segmentation} {Algorithms} in {3D} {Echocardiography},''
  \emph{IEEE Transactions on Medical Imaging}, vol.~35, no.~4, pp. 967--977,
  Apr. 2016.

\bibitem{tobon-gomezBenchmarkingFrameworkMyocardial2013}
C.~Tobon-Gomez \emph{et~al.}, ``Benchmarking framework for myocardial tracking
  and deformation algorithms: {An} open access database,'' \emph{Medical Image
  Analysis}, vol.~17, no.~6, pp. 632--648, Aug. 2013.

\bibitem{zhaoMITEADatasetMachine2022}
D.~Zhao \emph{et~al.}, ``\BIBforeignlanguage{eng}{{MITEA}: {A} dataset for
  machine learning segmentation of the left ventricle in {3D} echocardiography
  using subject-specific labels from cardiac magnetic resonance imaging},''
  \emph{\BIBforeignlanguage{eng}{Frontiers in Cardiovascular Medicine}},
  vol.~9, p. 1016703, 2022.

\bibitem{vukadinovicAutomatedInterpretable2D2025}
M.~Vukadinovic, H.~Ieki, Y.~Sahashi, D.~Ouyang, and B.~He, ``Automated
  {Interpretable} {2D} {Video} {Extraction} from {3D} {Echocardiography},''
  Nov. 2025, arXiv:2511.15946.

\bibitem{vukadinovicComprehensiveEchocardiogramEvaluation2026}
M.~Vukadinovic \emph{et~al.}, ``\BIBforeignlanguage{en}{Comprehensive
  echocardiogram evaluation with view primed vision language {AI}},''
  \emph{\BIBforeignlanguage{en}{Nature}}, vol. 650, no. 8103, pp. 970--977,
  Feb. 2026.

\bibitem{naidooRobustFinegrainedEchocardiographic2026}
P.~Naidoo \emph{et~al.}, ``\BIBforeignlanguage{en}{Robust fine-grained
  echocardiographic view classification with supervised contrastive
  learning},'' \emph{\BIBforeignlanguage{en}{Medical Image Analysis}}, vol.
  110, p. 104006, May 2026.

\bibitem{tan_efficientnet_2019}
M.~Tan and Q.~V. Le, ``{EfficientNet}: Rethinking model scaling for
  convolutional neural networks,'' in \emph{Proceedings of the 36th
  International Conference on Machine Learning}, ser. Proceedings of Machine
  Learning Research, vol.~97, 2019, pp. 6105--6114.

\bibitem{loshchilovDecoupledWeightDecay2019}
I.~Loshchilov and F.~Hutter, ``Decoupled {Weight} {Decay} {Regularization},''
  in \emph{International {Conference} on {Learning} {Representations}
  ({ICLR})}, 2019.

\bibitem{jordanMuonOptimizerHidden2024}
K.~Jordan \emph{et~al.}, ``Muon: {An} optimizer for hidden layers in neural
  networks,'' 2024.

\bibitem{langEAEASERecommendations2012}
R.~M. Lang \emph{et~al.}, ``\BIBforeignlanguage{en}{{EAE}/{ASE}
  {Recommendations} for {Image} {Acquisition} and {Display} {Using}
  {Three}-{Dimensional} {Echocardiography}},''
  \emph{\BIBforeignlanguage{en}{Journal of the American Society of
  Echocardiography}}, vol.~25, no.~1, pp. 3--46, Jan. 2012.

\bibitem{lyon2022ESCGuidelines2022}
A.~R. Lyon \emph{et~al.}, ``\BIBforeignlanguage{en}{2022 {ESC} {Guidelines} on
  cardio-oncology developed in collaboration with the {European} {Hematology}
  {Association} ({EHA}), the {European} {Society} for {Therapeutic} {Radiology}
  and {Oncology} ({ESTRO}) and the {International} {Cardio}-{Oncology}
  {Society} ({IC}-{OS})},'' \emph{\BIBforeignlanguage{en}{European Heart
  Journal}}, vol.~43, no.~41, pp. 4229--4361, Nov. 2022.

\bibitem{vahanian2021ESCEACTS2022}
A.~Vahanian \emph{et~al.}, ``2021 {ESC}/{EACTS} {Guidelines} for the management
  of valvular heart disease,'' \emph{EuroIntervention}, vol.~17, no.~14, pp.
  e1126--e1196, Feb. 2022.

\end{thebibliography}

\end{document}